\documentclass[journal=jctcce,manuscript=article,
               layout=traditional]{achemso}
\usepackage{graphicx}
\usepackage{dcolumn}
\usepackage{bm}
\usepackage{booktabs}
\usepackage{svg}
\usepackage{amsmath}
\usepackage{listings}
\usepackage[caption=false]{subfig}

\newcommand\hl[1]{\textcolor{blue}{#1}}
\let\hl\undefined
\newcommand{\hl}[1]{#1}

\author{William Dawson}
\email{william.dawson@riken.jp}
\author{Jens Domke}
\author{Takahito Nakajima}
\affiliation{RIKEN Center for Computational Science, Kobe, Japan}
\author{Katsuhisa Ozaki}
\affiliation{Shibaura Institute of Technology, Saitama, Japan}

\title{Reducing Numerical Precision Requirements in Quantum Chemistry Calculations}

\begin{document}

\begin{abstract}
The abundant demand for deep learning compute resources has created a renaissance in low precision hardware. Going forward, it will be essential for simulation software to run on this new generation of machines without sacrificing scientific fidelity. In this paper, we examine the precision requirements of a representative kernel from quantum chemistry calculations: calculation of the single particle density matrix from a given mean field Hamiltonian (i.e. Hartree-Fock or Density Functional Theory) represented in an LCAO basis. We find that double precision affords an unnecessarily high level of precision, leading to optimization opportunities. We show how an approximation built from an error-free matrix multiplication transformation can be used to potentially accelerate this kernel on future hardware. Our results provide a road map for adapting quantum chemistry software for the next generation of High Performance Computing platforms.
\end{abstract}

\maketitle

\section{Introduction}

In recent years, progress in the field of Artificial Intelligence has lead to an increase in demand for computing resources to perform deep learning~\cite{sevilla2023please}. This has significant implications for developments in computational quantum chemistry --- simulation software must be written to coexist on AI-centric software and hardware ecosystems. Of particular importance will be the ability to run quantum chemistry packages on low precision hardware such as NVIDIA's Tensor Cores or Google's TPUs (see, for example, Ref.~\cite{markidis2018nvidia} and Ref.~\cite{Lewis2022} respectively for their application to scientific problems). While targeting low precision hardware will increase the challenge of developing simulation software in the short term, it also presents new opportunities for co-designing specialized hardware optimal for solving scientific problems.

In this paper, we will systematically explore the floating point precision requirements necessary for a representative quantum chemistry kernel: calculation of the single particle density matrix represented in a linear combination of atomic orbitals (LCAO) basis set. We will find that double precision calculations provide an unnecessarily high level of precision. We will then propose the use of the error-free transformation for matrix multiplication developed by Ozaki and coworkers~\cite{ozaki2012error} (i.e. the Ozaki scheme) in combination with density matrix purification~\cite{palser1998canonical} to exploit low precision hardware while obtaining the required precision (and no more). 

\section{Background}

We will first review the key points of floating point calculations on modern hardware. Then we will discuss the history of low precision calculation algorithms in computational quantum chemistry. Subsequently we will introduce the target algorithm of density matrix purification and its relevance for low precision calculations (including recent promising work). Finally, we will present the Ozaki scheme for efficiently emulating higher precision matrix multiplication using low precision hardware.

\subsection{Floating Point Representations}

\begin{figure*}
\centering
\includegraphics[width=1\columnwidth]{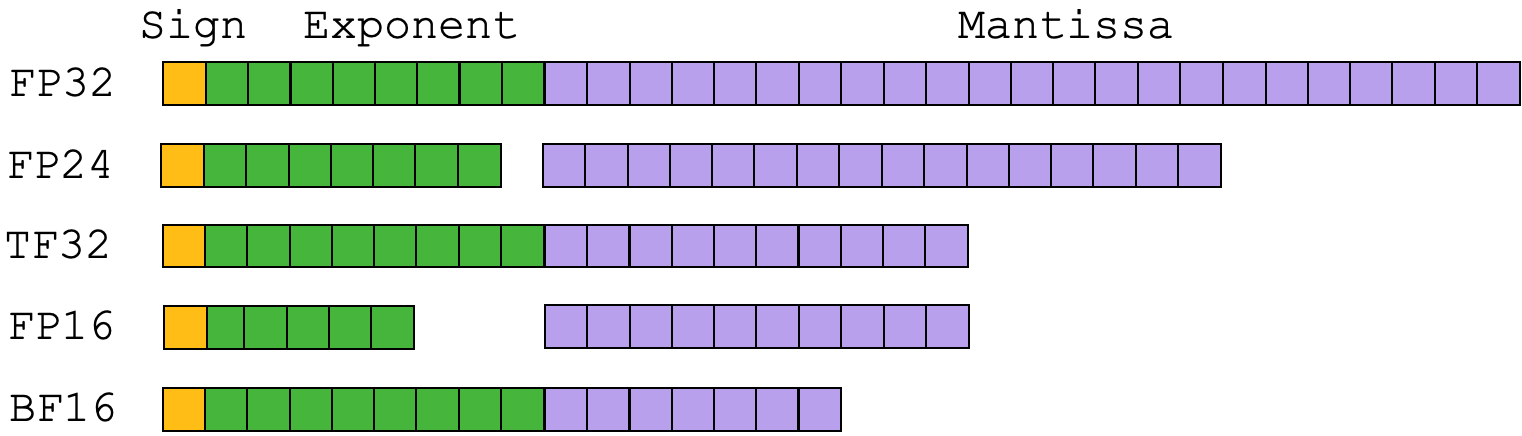}
\caption{Bitwise representation of various floating point types. FP32: single precision; FP24: AMD's FP24; TF32: NVIDIA's Tensor Float, FP16: Half Precision, BF16: BFLOAT16 (brain floating point).}
\label{fig:fp}
\end{figure*}

IEEE-754 floating point numbers are written in terms of a sign bit, exponent, and mantissa (Fig.~\ref{fig:fp}). For example, single precision (FP32) uses 8 bits for the exponent and 23 for the mantissa, whereas half precision (FP16) uses 5 and 10 respectively. Due to the implicit bit, these formats are able to effectively store 24 (FP32) and 11 (FP16) bits of precision in the mantissa. In quantum chemistry codes, the standard is to use double precision calculations (FP64), which uses 11 bits for the exponent and 52 for the mantissa. Recently, new floating point formats such as NVIDIA's TF32 (8, 10) or BFLOAT16 (8, 7) have been proposed specifically for machine learning applications where only a small mantissa is required. In this work, we will also consider AMD's FP24 (7, 16) format as an example of a type with a mantissa between the size of FP16 and FP32. 

\hl{The Tensor Cores (TC) available on modern GPUs are one way to exploit these low precision data types. Tensor Cores perform a fused multiply add (FMA) operation on $4\times 4$ matrices, with the accumulation potentially performed in a higher precision than the input matrices. For example, NVIDIA FP16 TCs will operate on FP16 matrices and accumulate in FP32. NVIDIA's TF32 TCs likewise take in TF32 matrices and accumulate in FP32. These specialized matrix operations can lead to substantial acceleration over standard floating point operations. In Tab.~\ref{tab:gpu_performance} we summarize the floating point performance for various precision combinations on different GPUs. We see that on an H100 SXM, using FP16 TCs instead of the FP64 TCs can potentially accelerate matrix multiplication by nearly 15x. The difference between FP64 and low precision operations is even larger when not using FP64 TCs (which are not always available). As demand for AI performance increases, we may anticipate this gap will increase further in the future.}

\begin{table*}[htbp]
\centering
\begin{tabular}{lccccc}
\toprule
\toprule
\text{GPU Model} & \text{FP64} & \text{FP64 TC} & TF32 TC & \text{FP16 TC} & \text{FP8 TC} \\ 
\midrule
\text{NVIDIA V100 SXM2~\cite{nvidia_v100_datasheet}} & 7.8 & \textendash &  \textendash & 125 & \textendash             \\ 
\text{NVIDIA A100 SXM~\cite{nvidia_a100_datasheet}}  & 9.7 & 19.5 & 156 & 312 & 624 \\ 
\text{NVIDIA H100 SXM~\cite{nvidia_h100_datasheet}}  & 34 & 67   & 494.5 & 990 & 1979      \\ 
\text{NVIDIA RTX 4500 ADA~\cite{gpu_list}}  & 0.619 & ---   & 79 & 159 & 317      \\ 
\text{NVIDIA L40S~\cite{gpu_list}}  & 1.4 & ---   & 183 & 366 & 733      \\ 
\text{AMD Instinct
MI300X~\cite{gpu_list}}  & 81.7 & 163 & 654 & 1307 & 2615      \\ 
\bottomrule
\bottomrule
\end{tabular}
\caption{\hl{Performance metrics, in TFLOPS, of various GPU models using different floating point representations. For the FP64 values, we report both the performance with and without FP64 Tensor Cores. We note that the performance of the Tensor Cores on newer machines can be increased by a factor of two when sparsity can be exploited.}}
\label{tab:gpu_performance}
\end{table*}

\subsection{Low Precision Quantum Chemistry Calculations}

There has been significant research on low precision computing for computational quantum chemistry, however it has primarily focused on single precision vs. double precision. Single precision for the analytic calculation of two-electron \hl{repulsion integrals} in a Gaussian basis set has been demonstrated to be an effective strategy to exploit GPUs~\cite{Yasuda2008, Ufimtsev2008, Luehr2011, Tornai2019}. Single precision can also be employed to accelerate semi-numerical methods~\cite{Laqua2021}. The single precision strategy has recently been applied to Slater type orbitals as well~\cite{dang2022numerical}. In materials science, single precision was shown to be a promising strategy to accelerate the computation of exact-exchange in codes using a planewave basis set~\cite{Vinson2020}. Single precision has also been employed to speed up the iterative eigenvalue solvers of materials codes\hl{~\cite{Tsuchida2012,10.1145/3295500.3357157, Das2022,10.1145/3581784.3627037,Woo2023,Khadatkar2023}}. Single precision calculations are particularly promising for many body perturbation theory methods. Early work demonstrated the reduced precision requirements of MP2 calculations~\cite{Vogt2008, Olivares-Amaya2010}, including when using the RI technique~\cite{Vysotskiy2011}. Single precision can also be applied to Coupled Cluster~\cite{DePrince2011, Pokhilko2018}, including time dependent variants~\cite{Wang2022}. Other targets of single precision optimizations include DMRG~\cite{Tian2022}, quantum transport calculations~\cite{10.1145/3295500.3357156}, and GW~\cite{Yu2022}.

A challenge the community has faced for developing low precision software has been to predict how relevant such optimizations will be to future architectures. For example, Yasuda showed that evaluation of the exchange and correlation functional could be accelerated using single precision~\cite{Yasuda20082}; however, a recent work~\cite{10.3389/fchem.2020.581058} explicitly rejected this strategy noting the gap in performance between double and single precision on GPUs has been closed. The earlier mentioned work on semi-numerical calculations~\cite{Laqua2021} justified its strategy by targeting lower-cost ``gaming GPUs'', where single precision continues to have higher performance. With recent developments in Artificial Intelligence, the pendulum has swung back towards the relevance of low (and more exotic) precision hardware. It is now crucial for the quantum chemistry community to establish clear precision requirements for their simulations. We note that even if double precision capable hardware remains the standard, studying the effects of low precision will still be potentially useful for reducing data transfer costs. 

\subsection{Density Matrix Purification}

In mean field quantum chemistry calculations, such as Kohn-Sham Density Functional Theory~\cite{hohenberg-inhomogeneous-1964, kohn-self_consistent-1965}, we need to compute the single particle density matrix from a given Hamiltonian. Limiting ourselves to the spin-restricted case, we expand the orbitals in some set of $M$ basis functions:
\begin{equation}
\psi_i(r) = \sum_j^M c_{ij}\phi_j(r).
\end{equation}
We in turn obtain matrix representations of our fundamental operators:
\begin{equation}
S_{ij} = <\phi_i|\hat{I}|\phi_j>,
\end{equation}
\begin{equation}
H_{ij} = <\phi_i|\hat{H}|\phi_j>,
\end{equation}
where $\hat{I}$ is the identity and $\hat{H}$ the Hamiltonian operator. This leads to the generalized eigenvalue problem:
\begin{equation}
H\psi_i = \lambda_i S \psi_i,
\label{eq:eig}
\end{equation}
from the solutions of which we can construct the single particle density matrix:
\begin{equation}
\hl{K_{ij} = \sum_a^M f_a c_{ia}c_{ja},}
\end{equation}
where $f$ is the occupation number (usually $2$ for occupied and $0$ for unoccupied orbitals for spin-restricted calculations of insulating systems).

In most implementations based on LCAOs, equation~\ref{eq:eig} is solved by invoking a dense eigenvalue solver, such as the ones available in LAPACK or ScaLAPACK. Unfortunately, these calculations have a computational cost that scales with the third power of the number of basis functions. To reduce this cost for application to large systems, many ``diagonalization free'' methods have been proposed~\cite{bowler-ON-2012}. One class of ``diagonalization free'' methods is based on the purification algorithm first proposed by McWeeny~\cite{RevModPhys.32.335}, which iteratively computes $K$ using the following recurrence relation:
\begin{align}
P_{0} = \frac{\lambda}{2}(\mu I - H) + \frac{1}{2}I, \\
P_{k + 1} = 3P_k^2 - 2P_k^3,
\end{align}
where $\mu$ is the chemical potential, $\lambda$ scales the spectrum of $H$ to be within the range $[0, 1]$, and $K=2P$. 

The power of such an approach is that the core computational kernel is matrix - matrix multiplication, which can readily exploit the underlying sparsity of $H$ and $K$ that exist for large insulating systems~\cite{Prodan2005} \hl{--- though whether sufficient sparsity exists to get a computational advantage depends on the system and choice of basis set}. A number of different purification methods exist (see, for example, the methods implemented in the NTPoly library~\cite{Dawson2018}); each employs different forms and orders of polynomials during the iterations. The second order trace correcting method of Niklasson~\cite{Niklasson2003} has the benefit of only needing to compute the square of a matrix, a point we will return to in Sec.~\ref{sec:markidis}.

Density matrix purification is not only useful for the case of extremely large systems where such sparsity exists. Since the purification algorithm has matrix - matrix multiplication as a bottleneck, it has the benefit of scaling better on supercomputers than eigenvalue solvers do. When developing a Hartree-Fock code for the Tianhe-2 supercomputer, Chow and coworkers utilized purification to substantially improved scalability~\cite{Chow2015,Chow2016}. Finkelsein and coworkers recently demonstrated how an algorithm similar to purification implemented on a GPU could outperform dense eigenvalue solvers even on a single GPU~\cite{Finkelstein2023}. 

Pederson and coworkers~\cite{Pederson2023} similarly have proposed using dense purification as a means of exploiting a cluster of Google TPUs. In their work, they use a mixed-precision scheme where early SCF iterations are performed in single precision and the final iterations in a software emulated double precision.  Around the same time, Finkelstein and coworkers performed a series of studies using the Tensor Cores available on NVIDIA GPUs~\cite{Finkelstein2021, Finkelstein20212}. They targeted single precision accuracy by taking advantage of the Tensor Core's ability to accumulate in single precision and employing the Markidis scheme~\cite{markidis2018nvidia}. The work was further extended to density functional perturbation theory~\cite{Finkelstein2022} \hl{and for computing the inverse square root of the overlap matrix~\cite{Habib2024}}. In our work here, we will go beyond these earlier studies to achieve the higher precision required for general application. 

\hl{We note that density matrix purification algorithms are only applicable to insulating systems. To address this, several different algorithms have been introduced that allow for the introduction of the electronic temperature while still having matrix multiplication as the bottleneck (see the Perspective of Aarons et al.~\cite{Aarons2016} or some more recent methods~\cite{Aarons2018,Mniszewski2019,Leamer2024}). In Supplementary Information I, we perform some experiments using the Fermi Operator Expansion~\cite{10.1103/PhysRevB.51.9455} as a representative finite temperature method to show that the findings of this paper transfer to the case of metallic systems.}

\subsection{Ozaki Scheme}
\label{sec:ozaki_scheme}

The Ozaki scheme~\cite{ozaki2012error} performs an error-free transformation of computing the product of two matrices into a summation of several matrix multiplications that can be performed without rounding error. Several implementations of the Ozaki scheme exist both to target higher than double precision~\cite{10.1007/978-3-030-43229-4_44, Ichimura2018} and for using low precision units like Tensor Cores~\cite{10.1007/978-3-030-50743-5_12, ootomo2023dgemm}. Remarkably, an implementation based on the INT8 Tensor Cores of an RTX A6000 GPU could outperform double precision cuBLAS by $>4\times$ without loss of accuracy. On future architectures where low precision arithmetic units further dominate FP64, this scheme would be even more potent.

\begin{figure*}
\centering
\includegraphics[width=1\columnwidth]{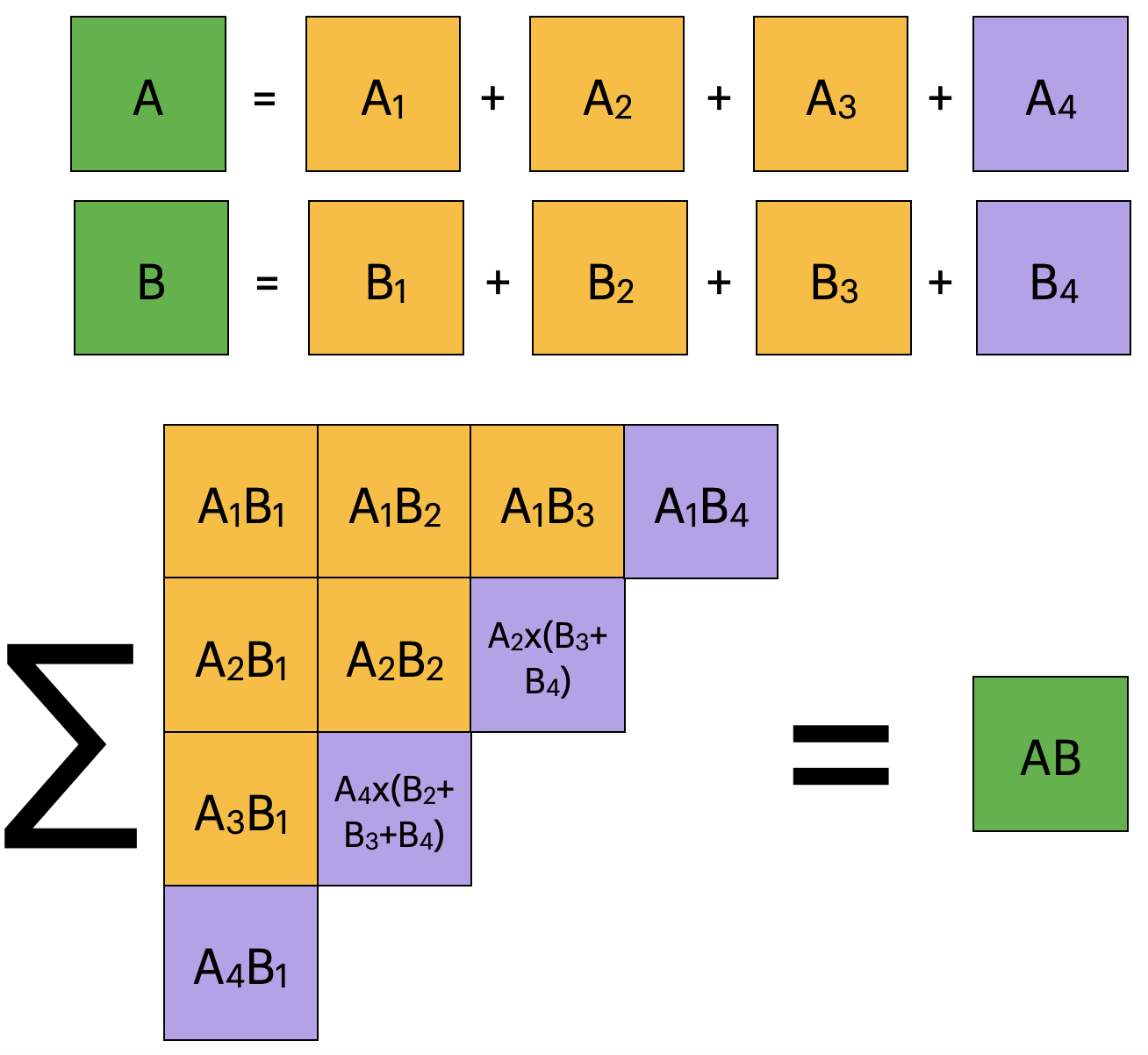}
\caption{An overview of the Ozaki scheme performed with four splits. The matrices A and B are split into three matrices that can be multiplied exactly in the target precision, and a remainder matrix. A sequence of multiplications is performed, which is summed up to form AB. The multiplications in the upper left can be performed with no error, whereas the final multiplications on the diagonal introduce the error from truncating the number of splits.}
\label{fig:ozaki_over}
\end{figure*}

The details of the Ozaki scheme have been presented in several previous publications, so we only briefly review the concept here (Fig.~\ref{fig:ozaki_over}). The scheme begins by splitting the input matrices into $S$ split matrices (of the same size as the original). This is done through a series of rounding and bit shifting operations so that each matrix can be represented exactly in the low precision representation (we include the improved version introduced later by Minamihata et al.~\cite{ozaki_conference}, see equation 3 in the paper of Mukunoki et al.~\cite{10.1145/3472456.3472493} for details). A further scaling operation is applied to maintain the exponent's range~\cite{10.1007/978-3-030-50743-5_12} (we note that in the scaling method of Mukunoki, only error free terms are considered, but in our implementation we extend the scaling to all terms).  We then compute the product of pairs $A_iB_j$, where $i+j \leq S+1$, as well as a set of remainder terms. Finally, the resulting matrices are summed up in the original precision. The accuracy of the final result depends on the number of splits; if both matrices are split $S$ times we require $2S$ times as much memory and $S(S+1)/2$ as many multiplications. 

\hl{The Ozaki scheme can be contrasted with the multiple component arithmetic approach, such as the popular double-double format~\cite{Hida2001}. In the multiple component approach, each individual entry of the matrix is expressed as the sum of lower precision types. Henry et al.~\cite{Henry2019} used this approach to leverage BFLOAT16 FMA units to achieve FP32 precision. The Ozaki scheme, on the other hand, splits the matrix into the sum of several lower precision matrices. Thus, the Ozaki scheme can be thought of as a Structure of Arrays approach, whereas multiple component arithmetic is an Array of Structures approach. The Structure of Arrays approach can exploit well matrix engines like Tensor Cores since the inner operation remains unchanged.}

\begin{figure*}
\centering
\includegraphics[width=1\columnwidth]{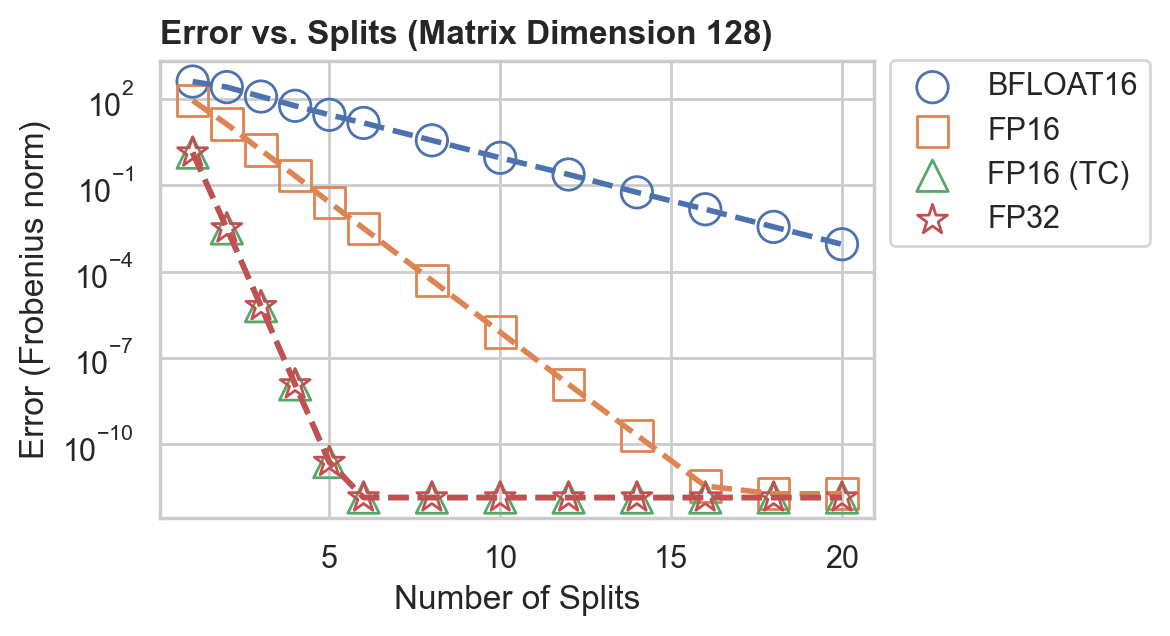}
\caption{Error in the multiplication of two random matrices as a function of the number of splits used in the Ozaki scheme. We examine the error for four different floating point types: BFLOAT16 (8, 7), FP16 (5, 10), FP16 with FP32 Accumulation (TC), and FP32 (8, 23).}
\label{fig:error}
\end{figure*}

In Fig.~\ref{fig:error}, we show some example calculations of the multiplication of two random matrices (FP64 elements distributed between [0, 1]) using different precisions and splits. Accuracy is measured as the Frobenius norm of the difference between the double precision reference result and the Ozaki scheme result. From this data, we see one crucial point about the Ozaki scheme: the floating point precision for accumulation determines the overall precision. Hence the Ozaki scheme is particularly effective at taking advantage of hardware like NVIDIA's Tensor Cores.

\section{Implementation Details}

For this study, we created two different implementations of density matrix purification: one based on NVIDIA's CUDA API for use on an NVIDIA RTX A6000 GPU and another using GNU MPFR to emulate low precision operations in software. MPFR will be particularly helpful to this study as it allows us to investigate arbitrarily defined floating point numbers~\cite{10.1145/1236463.1236468}. Thus we will use MPFR to explore precise precision requirements, and validate our results using the (substantially faster) CUDA implementation.

As input to our purification implementation, we use Hamiltonians coming from the PySCF~\cite{Sun2020} and BigDFT~\cite{Ratcliff2020} codes. We use HGH pseudopotentials~\cite{Willand2013} to remove the core electrons in BigDFT. BigDFT calculations are performed in the linear-scaling mode in order to produce a Hamiltonian in an LCAO basis set~\cite{Mohr2014}. PySCF calculations are done with the Polarization Consistent basis sets series~\cite{Jensen2014}. Calculations with BigDFT are performed with the PBE exchange-correlation functional~\cite{10.1103/PhysRevLett.77.3865} and for PySCF with B3LYP~\cite{Stephens1994}. 

For simplicity, we first transform equation~\ref{eq:eig} to the standard eigenvalue problem using the L{\"o}wdin method:
\hl{
\begin{align}
    \ddot{H} = S^{-\frac{1}{2}}HS^{-\frac{1}{2}}, \\
    K = S^{-\frac{1}{2}}\ddot{K}S^{-\frac{1}{2}}.
\end{align}
}
\hl{This transformation is performed by computing the inverse square root of the overlap matrix and transforming the Hamiltonian in double precision (though this could potentially be done with a diagonalization free method in low precision~\cite{Habib2024}). All analysis of errors is then performed using the orthogonalized Hamiltonian and density matrix.}

For our tests, we implement the Trace Resetting Fourth Order (TRS4) purification method~\cite{Niklasson2003}. We choose this method because it gives a clearer convergence signal than lower order methods. As a convergence criterion for the purification iterations, we use a change in the electronic energy, \hl{$Tr(\ddot{H}\ddot{K})$}, of $1\times 10^{-8}$ Hartree.
\hl{Due to the introduction of errors from simulating low precision, the purification algorithm may not be able to converge.}
For cases where convergence can't be achieved (due to low precision errors), the purification algorithm halts when the previous two energy values have increased, with an absolute change of less than $1\times10^{-2}$ Hartree (or on the 100th iteration). \hl{These failures usually happen when the final energy error is above $1\times10^{-8}$ Hartree, though there are some exceptions.}
The source code of our implementation is available online (gitlab.com/wddawson/ozp), \hl{including Jupyter notebooks for running the experiments (with extra data about convergence detection and exponent underflow)}.

\hl{\subsection{Error Evaluation Criteria}}

\hl{In this work, we will attempt to evaluate density matrix purification based on low precision arithmetic as a replacement for the dense eigenvalue solver used in LCAO codes. Of particular importance for such a replacement is that it should not impact convergence of the self-consistent field (SCF) iterations due to numerical noise. If a reliable ground state solution is found, properties may then be computed using a different precision. In Supplementary Information II, we demonstrate this for the case of the nuclear gradients.}

\hl{To monitor SCF convergence, codes based on an LCAO basis set use a variety of different measures. In NWChem 7.2.2, convergence is based on three different criteria. First, there is the root mean square deviation (RMSD) of changes in the density matrix. Second, there is the change in the electronic energy, $Tr(\ddot{H}\ddot{K})$, between iterations. Third, the norm of the change in the commutator DIIS error vector is used~\cite{Pulay1982}:}
\begin{equation}
||\ddot{H}\ddot{K} - \ddot{K}\ddot{H}||_F.
\end{equation}
\hl{The default convergence criteria for these measures when performing DFT calculations are $1\times10^{-5}$ (RMSD), $1\times10^{-6}$ Hartree (energy), and $5\times10^{-4}$ (DIIS), respectively. Some other codes use stricter convergence criteria by default. In Gaussian 16, the default cutoff for the RMSD is $1\times10^{-8}$. In PySCF 2.7.0, the convergence is based on the change in the energy being below $1\times10^{-9}$.}

\hl{Here we target to achieve a precision in the RMSD, energy, and DIIS error vector at about two orders of magnitude below each of the default settings for NWChem. An error above this in the density matrix may make it difficult to test for convergence as numerical precision leads to fluctuations. An error in the energy may also lead to unacceptable noise in properties like energy differences. If the Hamiltonian does not commute with the density matrix, the optimization procedure will take steps related purely to numerical noise, which could severely hamper convergence. Approximately two orders of magnitude would also ensure that tighter convergence thresholds can be used when needed, such as when computing numerical gradients to build the Hessian matrix.}

\section{Numerical Experiments}

We will now perform numerical experiments to understand what level of precision will be required for practical calculations. Our first experiments will establish the fact that double precision provides more than the necessary precision, opening up the way for approximate calculations. We will then investigate the Ozaki scheme as a means of achieving the target precision. We will further apply the Ozaki scheme to larger datasets of matrices to verify the robustness of our findings. We will also compare these results to the competing Markidis method. Finally, based on our findings, we will consider how current implementations of purification could benefit from a mixed precision scheme. We note that for all calculations, low precision is only used for the matrix multiplications, and double precision everywhere else.

\begin{figure*}
    \subfloat{
        \includegraphics[width=0.3\textwidth]{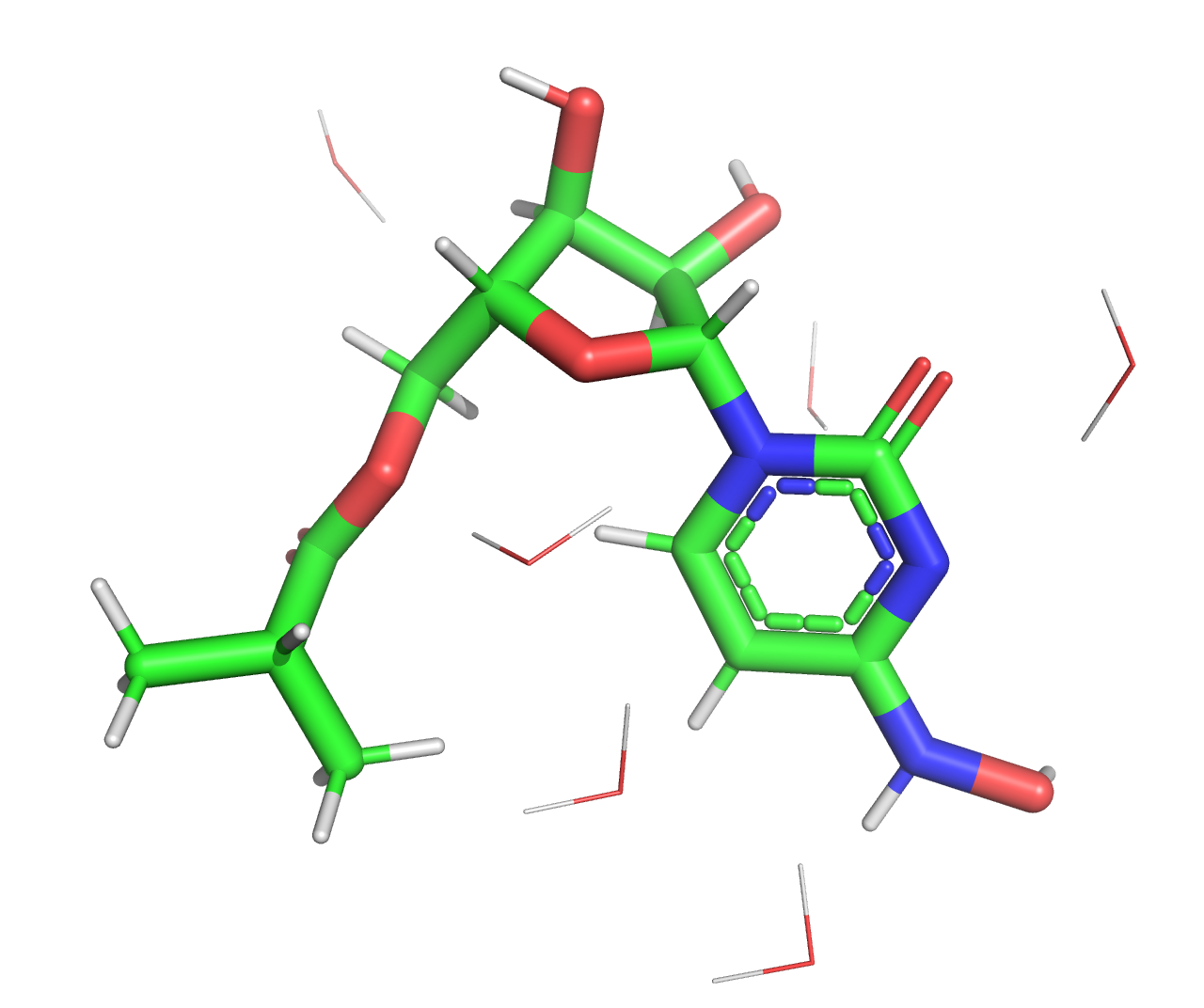}
    }
    \hfill
    \subfloat{
        \includegraphics[width=0.3\textwidth]{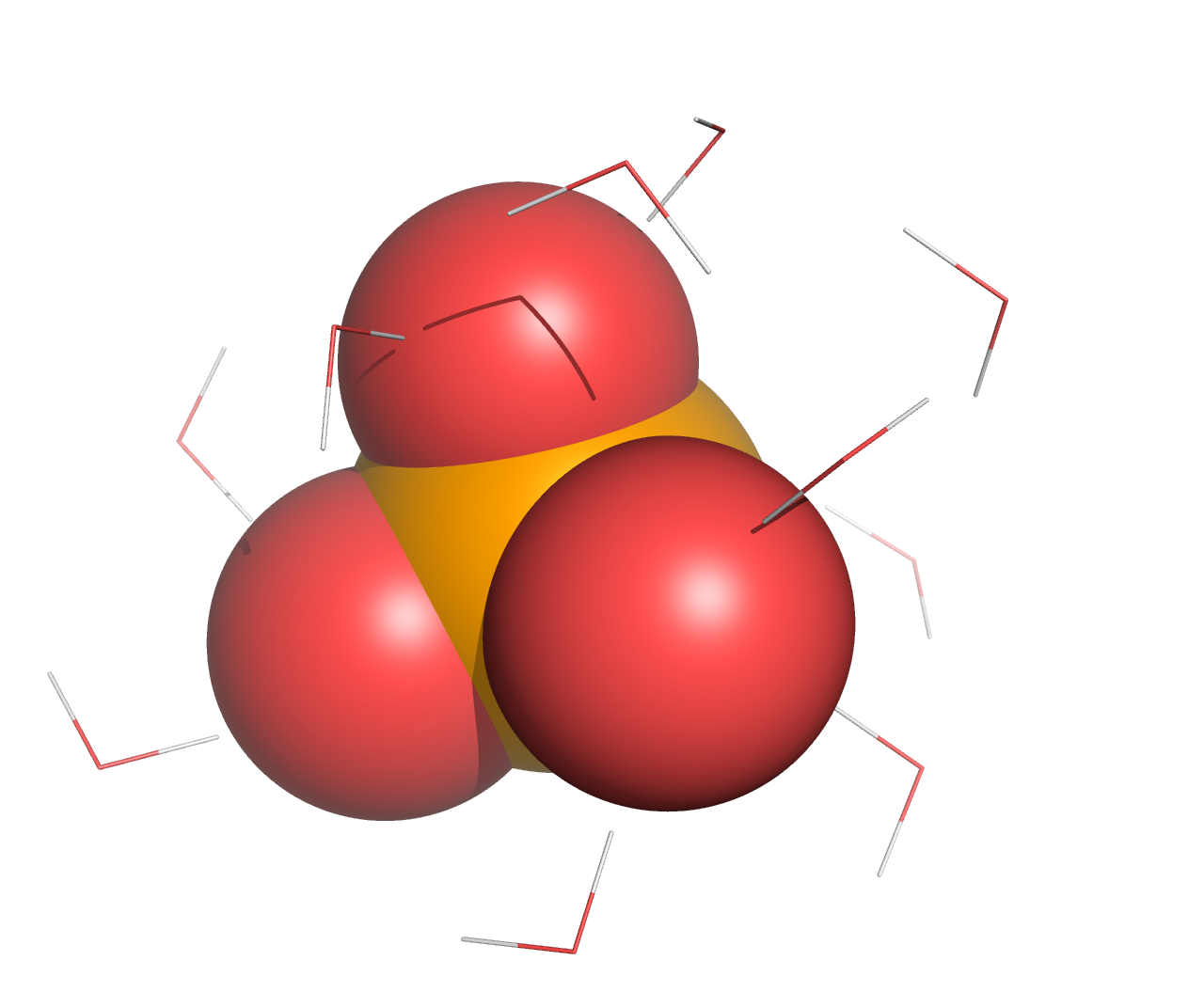}
    }
    \hfill
    \subfloat{
        \includegraphics[width=0.3\textwidth]{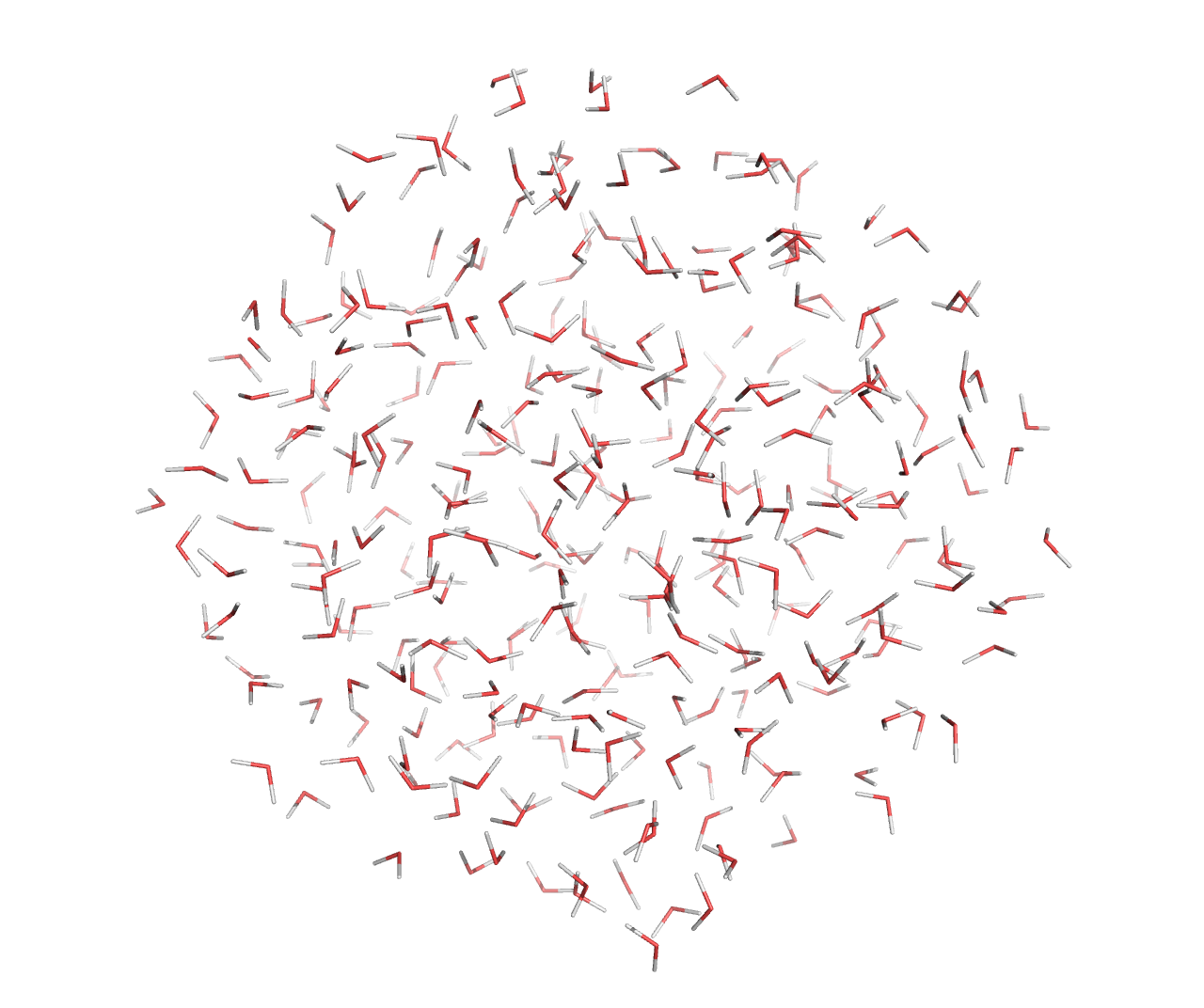}
    }
\caption{From left to right, the main systems used in this text: Molnupiravir, Selenite, and Water 237.} 
\label{fig:sys}
\end{figure*}

As test cases, we will first use the Hamiltonian coming from a BigDFT calculation of a Molnupiravir molecule bound to six water molecules. We will then expand the dataset to include a selenite ion surrounded by 10 water molecules calculated with different basis sets in PySCF. We also will include water clusters of different sizes and bulk silicon computed with BigDFT. We finally analyze a large water cluster (573 molecules) computed with B3LYP/PCSEG using NTChem~\cite{Nakajima2015, Dawson2023} (due to the system size). The main systems used in this paper are show in Fig~\ref{fig:sys}.

\subsection{Precision Requirements Sweep}
\label{sec:sweep}

\begin{figure*}
\centering
\includegraphics[width=1\columnwidth]{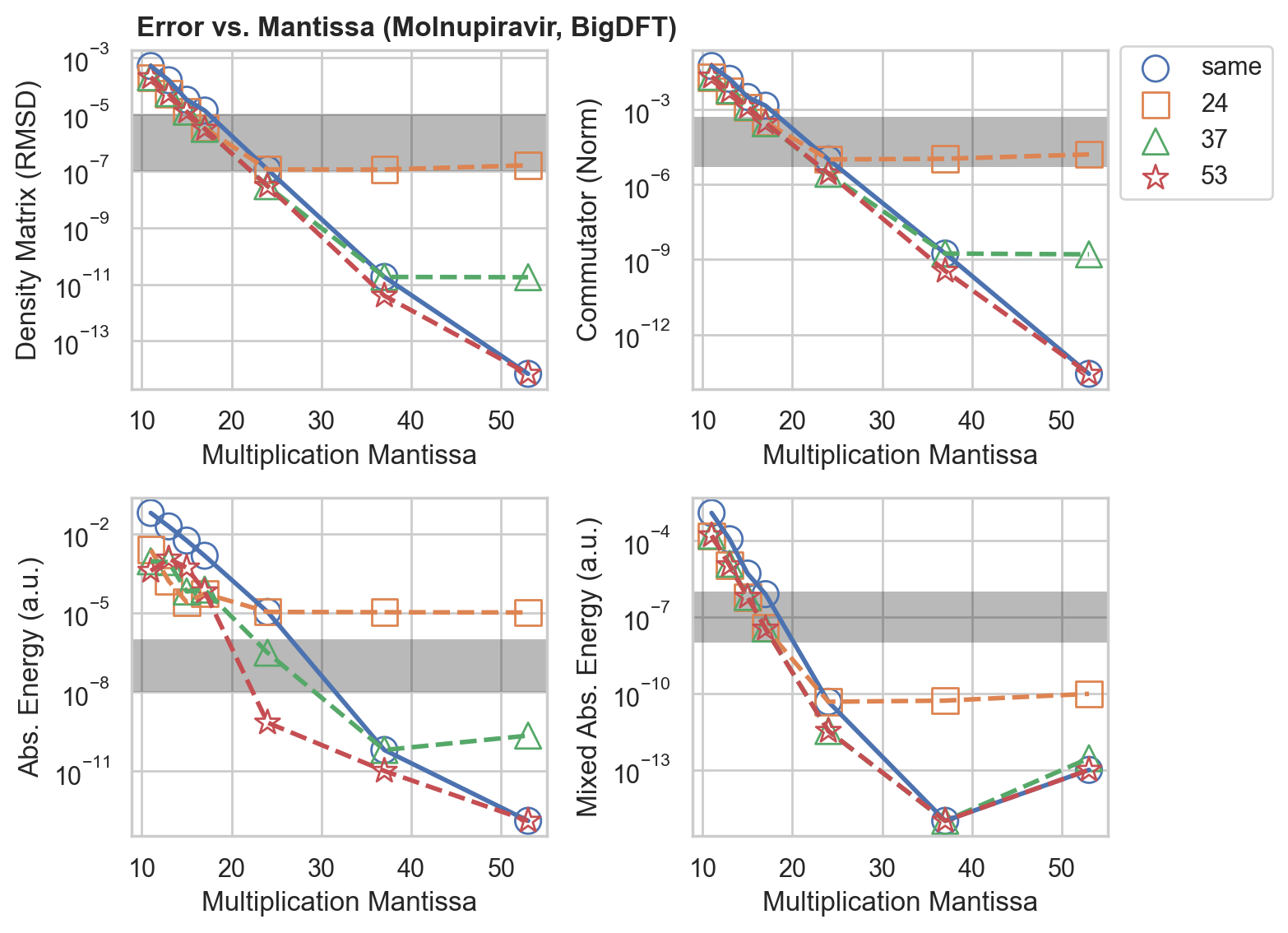}
\caption{\hl{Errors when using purification vs. various mantissa sizes (including the implicit bit) for multiplication and accumulation applied to the Molnupiravir / BigDFT system (matrix dimension $147\times 147$). Different lines represent different values of accumulation precision and the horizontal axis represents the precision used for multiplication. The ``same'' line uses the same precision for multiplication and precision. The gray box extends from the NWChem convergence cutoffs down to two orders of magnitude lower. ``Mixed Abs. Energy'' labels the energy error after applying one McWeeny step in full double precision.}}
\label{fig:sweep}
\end{figure*}

In order to understand the precision requirements for computing the density matrix, we perform density matrix purification using different custom floating point types implemented using MPFR on the Molnupiravir system. For this experiment, we assume \hl{double precision range for the exponent}. In Fig.~\ref{fig:sweep}, we plot the error of the resulting density matrix\hl{, commutator, and energy} compared to a reference result computed with diagonalization in double precision. We vary the effective precision of the mantissa used for multiplication from 11 (half) to 53 (double). For each calculation, we use one of three fixed accumulation mantissa values (24, 37, 53) or the ``same'' precision as multiplication.

\hl{From this data, we see that single precision is not sufficient to achieve a converged result. The errors in the density matrix and commutator are not quite below the two orders of magnitude range set by our NWChem target (much less the codes with stricter criteria). Interestingly, we observed that the error in the energy was significantly worse compared to the other measures. Following the suggestion Finkelstein et al.~\cite{Finkelstein2021}, we investigated a mixed precision approach, where following the the convergence of the purification algorithm we perform one purification step using full double precision. Specifically, in our case we take one McWeeny step:}
\begin{equation}
    P = 3P^2 - 2P^3.
\end{equation}
\hl{We found that applying this iterative refinement substantially improved the energy values, though this improvement did not extend to other measures. For example, with a single precision mantissa the absolute energy error improved from $1.08\times 10^{-5}$ to $4.67\times 10^{-11}$; however, the RMSD and commutator norm only changed from $1.13\times 10^{-7}$ to $1.09\times 10^{-7}$ and $1.00\times 10^{-5}$ to $9.99\times 10^{-6}$, respectively. We further found that the magnitude of the error in the RMSD for any given multiplication in the iterations remains roughly the same, so there would be little benefit to adjusting the precision across iterations.  Overall, we recommend the iterative refinement strategy when precision in the energy is important (such as the final SCF iteration).}

\hl{Another interesting observation is that a substantial amount of error in single precision comes from the accumulation phase, not the multiplication step. Importantly, we found that 37 bits of effective precision is sufficient for accumulation; for a single precision multiplication with 37 bits of accumulation, the errors are nearly indistinguishable from double precision accumulation. It is thus possible to obtain a high precision result while storing (and hence communicating) intermediate matrices in single precision if they are multiplied in a higher precision (a point we will return to in Sec.~\ref{sec:comm}). Overall, we find that the double precision commonly used for quantum chemistry calculations essentially provides an unnecessarily high level of precision for computing the single particle density matrix.}

\begin{figure*}
\centering
\includegraphics[width=1\columnwidth]{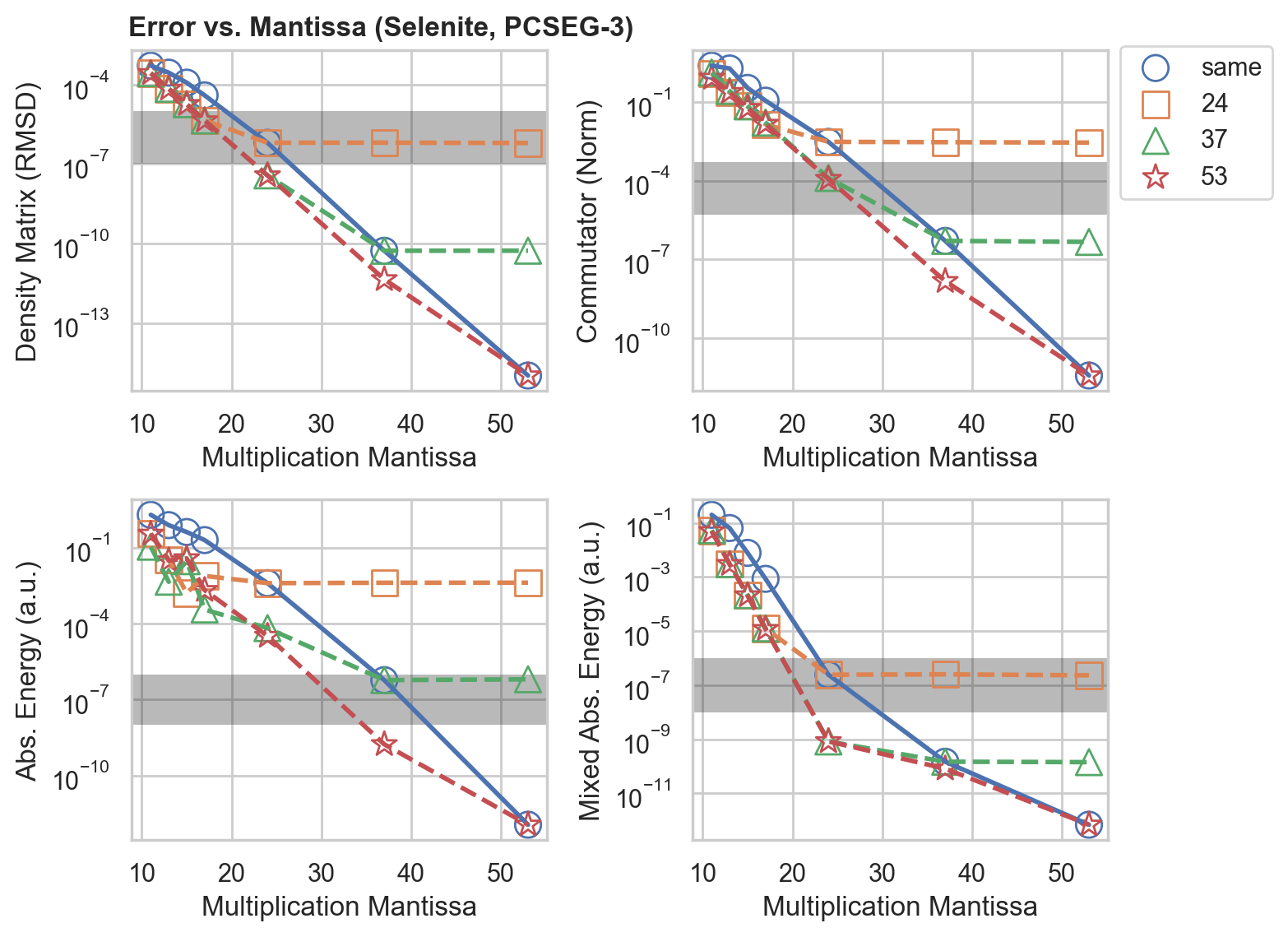}
\caption{\hl{Errors when using purification vs. various mantissa sizes (including the implicit bit) for multiplication and accumulation applied to the selenite / PCSEG-3 system (matrix dimension $1508\times 1508$). Different lines represent different values of accumulation precision and the horizontal axis represents the precision used for multiplication. The ``same'' line uses the same precision for multiplication and precision. The gray box extends from the NWChem convergence cutoffs down to two orders of magnitude lower. ``Mixed Abs. Energy'' labels the energy error after applying one McWeeny step in full double precision.}} 
\label{fig:sweep_c}
\end{figure*}

\hl{We now repeat this experiment using the selenite system with the PCSEG-3 basis set (Fig.~\ref{fig:sweep_c}). The overall pattern is very similar, however the errors are shifted up compared to the BigDFT calculation. We thus conclude that the quadruple-$\zeta$ basis set represents more challenging numerical conditions and hence requires larger floating point data types. Even in this case though, FP64 provides far more precision than is required. If refinement is applied to the energy values, the 37 effective bit mantissa provides a sufficient amount of precision, and may provide acceptable results when used to accumulate the multiplication of single precision matrices. Overall, we find that low precision algorithms show promise for these problems, particularly if they can efficiently exploit the fact that less than full double precision is required.}

\subsection{Ozaki Scheme Application}

\begin{figure*}
\centering
\includegraphics[width=1\columnwidth]{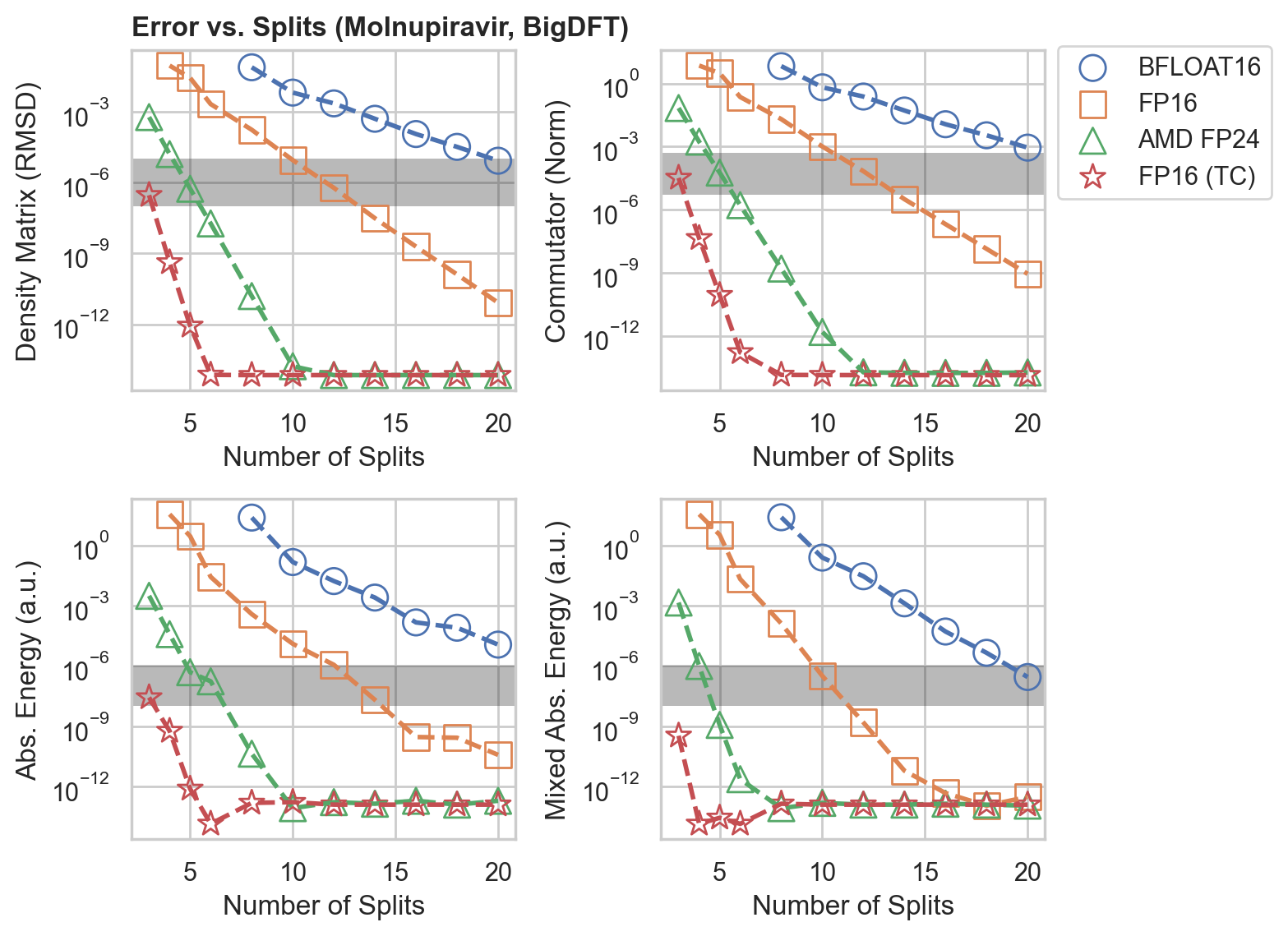}
\caption{\hl{Errors when using purification vs. the number of splits used in the Ozaki scheme applied to the Molnupiravir / BigDFT system (matrix dimension $147\times 147$). We examine the error for four different floating point types: BFLOAT16 (8, 7), FP16 (5, 10), AMD FP24 (7, 16), and FP16 with FP32 (8, 23) Accumulation (TC). The gray box extends from the NWChem convergence cutoffs down to two orders of magnitude lower. ``Mixed Abs. Energy'' labels the energy error after applying one McWeeny step in full double precision.}}
\label{fig:sweeperror}
\end{figure*}

Despite this promising finding, it is unlikely that future hardware will offer implementations of floating point types with effective 37 bit mantissas. Instead, we propose using the Ozaki scheme with a reduced number of splits to take advantage of the reduced precision needs. In Fig.~\ref{fig:sweeperror}, we plot the errors as a function of the number of splits for various floating point representations. When using NVIDIA's Tensor Cores (FP16 with FP32 accumulation), \hl{three to four splits (6 -- 10 multiplications)} is sufficient for a converged result. By contrast, according to Fig.~\ref{fig:error} to fully reproduce a double precision result on random matrices required 6 splits / 21 multiplications. Hence, the Ozaki scheme allows for approximately a factor of two savings in computational cost due to the lower precision requirements for quantum chemistry application. \hl{These multiplication counts can be compared with Tab.~\ref{tab:gpu_performance}, where we see that FP16 TCs can have above 15 times the performance of FP64 TCs, and over a factor of 100 more performance for standard double precision operations}.

In all cases, the scaling scheme seamlessly handles the exponent, and hence the precision comes down to the size of the mantissa. Using plain FP16, \hl{14 to 16 splits (105 -- 136 multiplications)} is required. This is a substantial increase in the number of multiplications compared to the Tensor Core version, however future hardware may not offer FP16 TC (especially non-NVIDIA chips), with different performance ratios. For a fully converged double precision result on random matrices, 20 splits / 210 multiplications was required, leading again to an approximate factor of two savings. We compare as well the intermediate sized mantissa of AMD's FP24, and find that it converges around 7 splits / 28 multiplications. For this system, the TRS4 algorithm required 26 calls to the Ozaki scheme multiplication routine to converge. Hence, when using \hl{FP16 TCs} an overall speedup would require a time to solution ratio of \hl{156 -- 260} multiplications to one double precision diagonalization. For pure FP16, the requirement increases to \hl{2756 -- 3536} multiplications. 

\subsection{Potential to Exploit Sparsity}
Depending on the underlying distribution of matrix values, the splitting procedure may introduce a significant amount of sparsity. \hl{Ichimura et al.~\cite{Ichimura2018} proposed to exploit the Ozaki scheme induced sparsity on a CPU by converting matrices with a sparsity level above 90\% to the ELL format, and performing sparse matrix - dense matrix multiplication. Sparsity can be more directly exploited to accelerate calculations on Tensor Cores:} on an A100 GPU, the FP16 TC performance increases from 312 TFLOPS to 624 TFLOPS for sparse matrices. \hl{This sparsity feature requires a special sparsity structure --- for any given row / column of four elements, at least two of them must be zero. }

We examined the final multiplication performed for the Molnupiravir system by the purification algorithm when using the Tensor Core representation (FP16 with FP32 accumulation) and \hl{four} splits \hl{to see how frequently this sparsity pattern appeared}. An entry of the split matrix (after scaling) was considered to be zero if it fell below the smallest value representable by double precision. The first split of both matrices have a number of non-zero \hl{blocks} below \hl{31\%}, however the next split is above \hl{95\%}, and subsequent splits are nearly fully dense. For the pure FP16 implementation with \hl{14} splits, the matrices are substantially more sparse: the number of non-zeros \hl{blocks} is below \hl{1\%} for the first split, \hl{3\%} for the second split, and \hl{9\%} for the third split. By the \hl{seventh} split the sparsity of both matrices was above \hl{59\%} (and for all subsequent splits). Thus we anticipate that some sparsity may be exploited in future work, though it would be limited to the first set of splits.


\subsection{Basis Set and Size Effects}

We now examine some larger datasets to determine the robustness of these previous results. For this analysis, we employ our CUDA implementation of the Ozaki scheme and focus on the FP16 Tensor Core version with three, four, five, and six splits. For these calculations, we first focus on the selenite system. We examine basis sets with up to quadruple-$\zeta$ quality as well as augmented versions. As matrices coming from Gaussian basis set calculations have worse conditioning (overlap matrix) and larger spectral widths (Hamiltonian) than the in-situ optimized support functions of BigDFT, this test examines the numerical stability of our findings. The selenite ion is an anion (-2) and selenium is a fourth row element, making it a suitable test case for larger basis sets.

\begin{figure*}
\centering
\includegraphics[width=1\columnwidth]{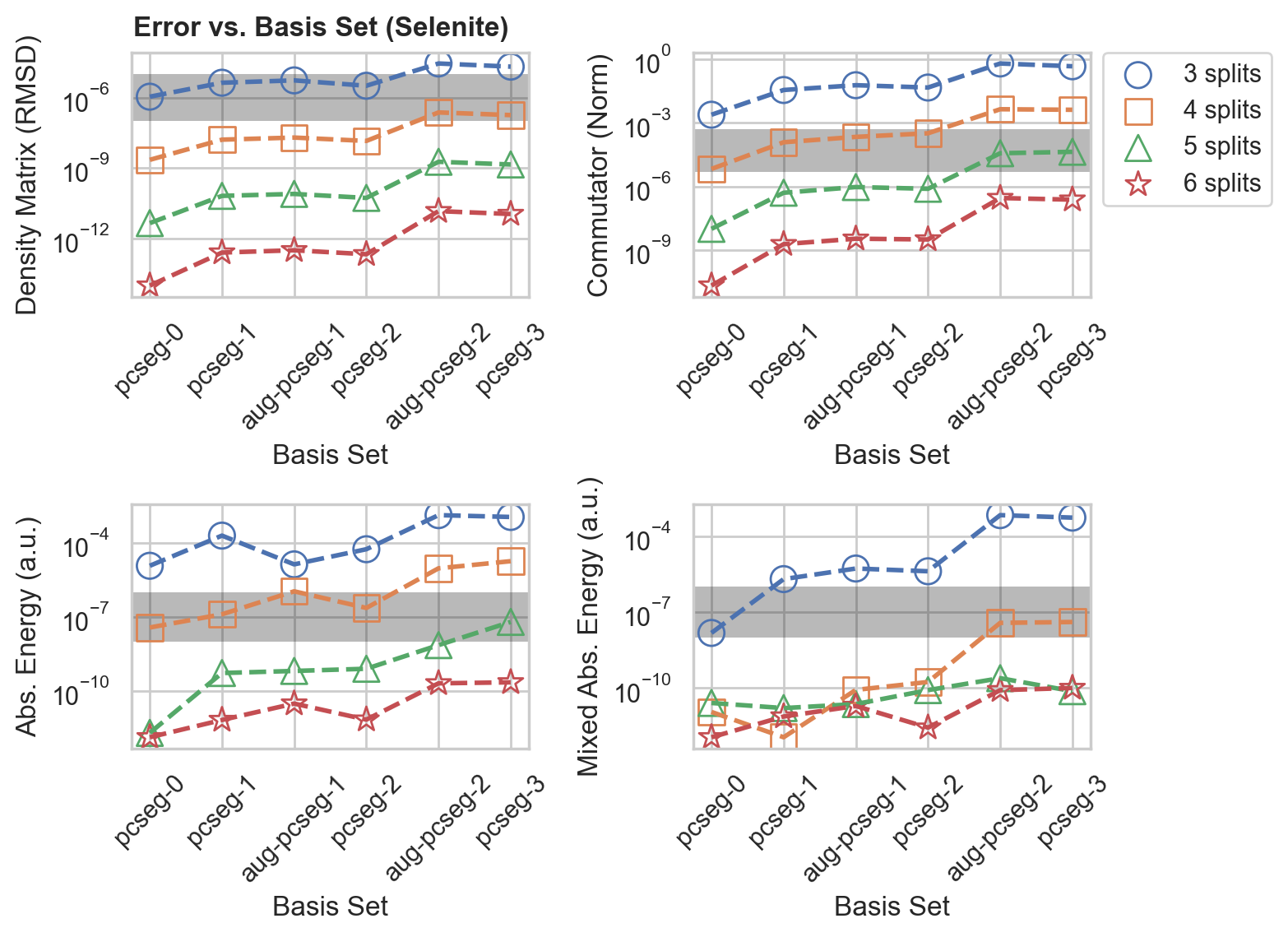}
\caption{\hl{Impact of the basis set on the number of splits required in the Ozaki scheme for the selenite in water system. Matrix sizes are: $179\times 179$ (PCSEG-0), $309\times 309$ (PCSEG-1), $515\times 515$ (AUG-PCSEG-1), $713\times 713$ (PCSEG-2), $1117\times 1117$ (AUG-PCSEG-2), and $1508\times 1508$ (PCSEG-3). All calculations are performed with FP16 Tensor Cores. The gray box extends from the NWChem convergence cutoffs down to two orders of magnitude lower. ``Mixed Abs. Energy'' labels the energy error after applying one McWeeny step in full double precision.}}
\label{fig:basis}
\end{figure*}

In Fig.~\ref{fig:basis} we plot the error for each of the basis sets with a given number of splits. We observe that for the larger basis sets, the required number of splits increases from four to five \hl{--- particularly for the errors in the commutator and energy}. This is consistent with our previous results where the difference between four and five splits was small, but not negligible. Overall, our earlier findings transfer well to even challenging numerical conditions. Thus, purification with the Ozaki scheme should be applicable to routinely performed quantum chemistry calculations, and not just specialized approximate schemes.

\begin{figure*}
\centering
\includegraphics[width=1\columnwidth]{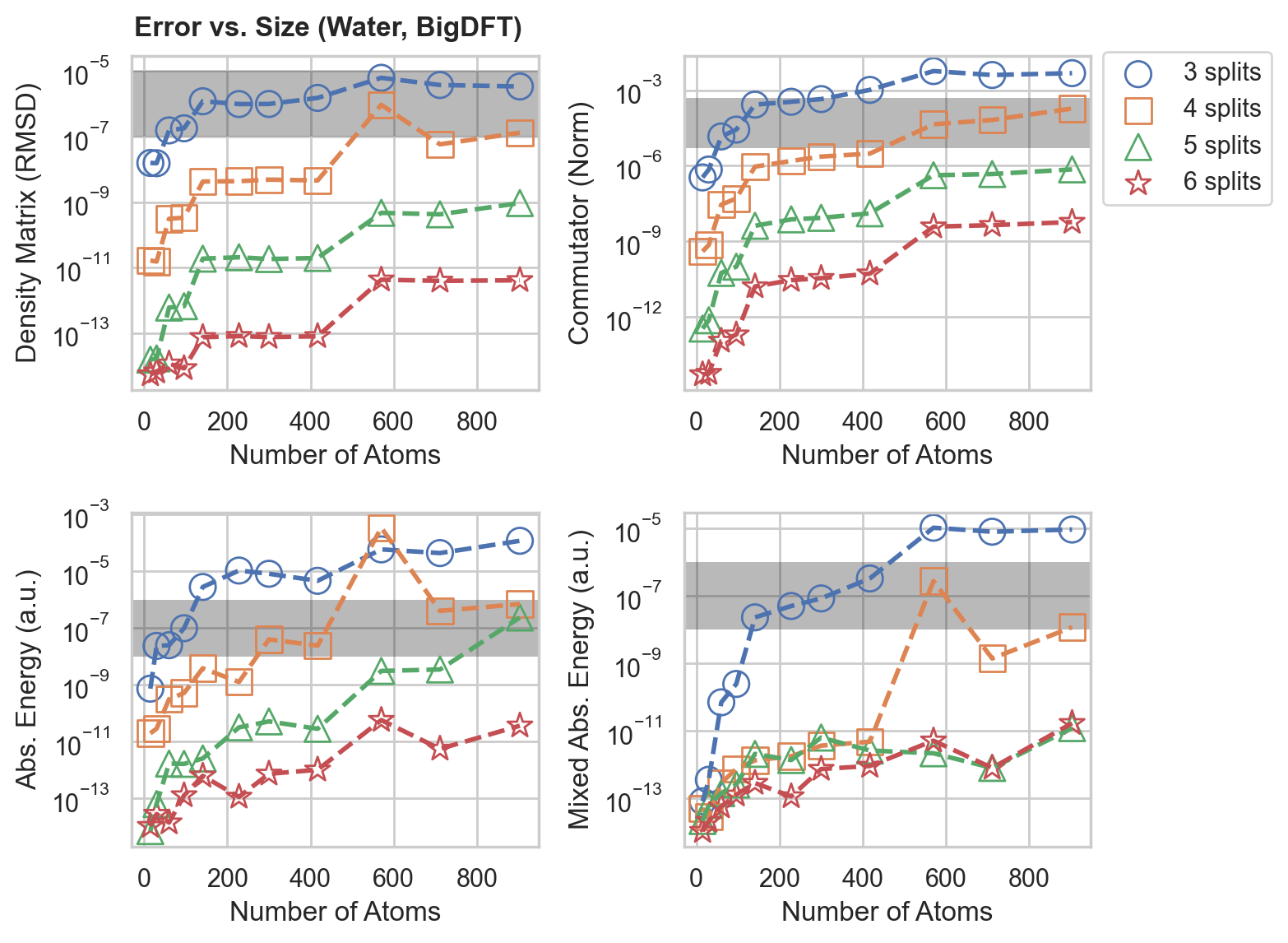}
\caption{\hl{Impact of the system size on the number of splits required in the Ozaki scheme for the water cluster systems. BigDFT uses 6 basis functions per water molecule, so that the smallest matrix is of dimension $30\times 30$ and the largest $1806\times 1806$. All calculations are performed with FP16 Tensor Cores. The gray box extends from the NWChem convergence cutoffs down to two orders of magnitude lower. ``Mixed Abs. Energy'' labels the energy error after applying one McWeeny step in full double precision.}}
\label{fig:size}
\end{figure*}

We also investigate the number of splits required as a function of system size using water clusters of increasing size computed with BigDFT. This analysis will allow us to separate the effect of basis set conditioning with matrix size. For the water clusters, the largest matrix is of dimension $1806\times 1806$, and for the selenite systems it was $1508\times 1508$ (PCSEG-3). The errors with respect to system size are plotted in Fig.~\ref{fig:size}. We find a modest growth in the number of splits required with system size. The error with five splits is similar for the largest water cluster as for the selenite system with the largest basis set, \hl{with the commutator and energy errors being about an order of magnitude lower}. Thus, we conclude that five splits is a sufficient recommendation, though automated schemes may improve the usability of the Ozaki scheme.

\subsection{Comparison with the Markidis Method}
\label{sec:markidis}
In the previous work of Finkelstein et al. that performed density matrix purification on NVIDIA Tensor Core units~\cite{Finkelstein2021, Finkelstein20212, Finkelstein2022}, the precision of the result was improved using the method of Markidis~\cite{markidis2018nvidia}. Here, the matrix $X$ is split into a lower and high precision part:
\begin{align}
X^0 = FP16[X], \\
X^1 = FP16[X - X^0],
\end{align}
after which the product $AB$ can be approximated as:
\begin{equation}
AB = FP32[A^0B^0 + A^0B^1 + A^1B^0 + A^1B^1].
\end{equation}
In the Markidis method, we can include up to four terms, each refining the precision of the result. In Tab.~\ref{tab:markidis} we show the precision and the number of purification iterations required (which may increase due to low precision) for a given number of terms for various systems. We caution that the number of iterations may be made more uniform by a more sophisticated convergence test~\cite{Kruchinina2016}.

\begin{table*}[htbp]
    \scriptsize
    \centering
    \begin{tabular}{lccccccc}
        \toprule
        \toprule
        System & 1 Term & 2 Terms & 3 Terms & 4 Terms & 4-MPFR & Ozaki (5) & FP64 \\
        \midrule
        \multicolumn{8}{c}{RMSD Errors} \\
        \midrule
        Molnupiravir & $2.2 \times 10^{-4}$ & $1.4 \times 10^{-4}$ & $5.1 \times 10^{-7}$ & $4.4 \times 10^{-7}$ & $2.1 \times 10^{-7}$ & $8.5 \times 10^{-13}$ & $7.0 \times 10^{-15}$ \\
        Water 237 & $6.1 \times 10^{-5}$ & $3.7 \times 10^{-5}$ & $2.0 \times 10^{-7}$ & $1.8 \times 10^{-7}$ & $1.1 \times 10^{-7}$ & $4.2 \times 10^{-10}$ & $8.1 \times 10^{-15}$ \\
        Selenite/A & $5.6 \times 10^{-4}$ & $3.2 \times 10^{-4}$ & $1.1 \times 10^{-6}$ & $1.1 \times 10^{-6}$ & $5.8 \times 10^{-7}$ & $6.5 \times 10^{-11}$ & $1.1 \times 10^{-15}$ \\
        Selenite/B & $3.8 \times 10^{-4}$ & $2.3 \times 10^{-4}$ & $1.2 \times 10^{-6}$ & $1.3 \times 10^{-6}$ & $8.6 \times 10^{-7}$ & $7.8 \times 10^{-11}$ & $1.3 \times 10^{-15}$ \\
        Selenite/C & $2.7 \times 10^{-4}$ & $1.6 \times 10^{-4}$ & $2.8 \times 10^{-6}$ & $2.9 \times 10^{-6}$ & $8.0 \times 10^{-7}$ & $1.4 \times 10^{-9}$ & $1.0 \times 10^{-15}$ \\
        \midrule
        \multicolumn{8}{c}{Commutator Errors} \\
        \midrule
        Molnupiravir & $2.2 \times 10^{-2}$ & $1.3 \times 10^{-2}$ & $5.5 \times 10^{-5}$ & $4.6 \times 10^{-5}$ & $2.3 \times 10^{-5}$ & $8.6 \times 10^{-11}$ & $2.8 \times 10^{-14}$ \\
        Water 237 & $6.5 \times 10^{-2}$ & $4.1 \times 10^{-2}$ & $2.2 \times 10^{-4}$ & $1.9 \times 10^{-4}$ & $1.1 \times 10^{-4}$ & $4.7 \times 10^{-7}$ & $2.9 \times 10^{-13}$ \\
        Selenite/A & $4.0 \times 10^{-1}$ & $1.5 \times 10^{-1}$ & $4.2 \times 10^{-3}$ & $4.5 \times 10^{-3}$ & $4.3 \times 10^{-3}$ & $5.1 \times 10^{-7}$ & $4.1 \times 10^{-13}$ \\
        Selenite/B & $4.1 \times 10^{-1}$ & $1.6 \times 10^{-1}$ & $5.2 \times 10^{-3}$ & $5.7 \times 10^{-3}$ & $4.6 \times 10^{-3}$ & $9.5 \times 10^{-7}$ & $1.4 \times 10^{-12}$ \\
        Selenite/C & $1.3 \times 10^{0}$ & $6.6 \times 10^{-1}$ & $1.4 \times 10^{-2}$ & $1.6 \times 10^{-2}$ & $1.2 \times 10^{-2}$ & $4.3 \times 10^{-5}$ & $3.8 \times 10^{-12}$ \\
        \midrule
        \multicolumn{8}{c}{Abs. Energy Errors} \\
        \midrule
        Molnupiravir & $2.6 \times 10^{-3}$ & $1.9 \times 10^{-3}$ & $8.7 \times 10^{-5}$ & $8.5 \times 10^{-5}$ & $1.2 \times 10^{-5}$ & $4.8 \times 10^{-13}$ & $1.1 \times 10^{-13}$ \\
        Water 237 & $1.9 \times 10^{-2}$ & $2.2 \times 10^{-5}$ & $1.2 \times 10^{-3}$ & $1.2 \times 10^{-3}$ & $3.6 \times 10^{-4}$ & $3.5 \times 10^{-9}$ & $1.7 \times 10^{-12}$ \\
        Selenite/A & $1.3 \times 10^{-1}$ & $1.4 \times 10^{-3}$ & $4.3 \times 10^{-3}$ & $4.3 \times 10^{-3}$ & $1.0 \times 10^{-3}$ & $5.4 \times 10^{-10}$ & $0$ \\
        Selenite/B & $5.8 \times 10^{-2}$ & $8.7 \times 10^{-2}$ & $8.5 \times 10^{-3}$ & $8.4 \times 10^{-3}$ & $2.2 \times 10^{-3}$ & $6.5 \times 10^{-10}$ & $4.5 \times 10^{-13}$ \\
        Selenite/C & $3.5 \times 10^{-1}$ & $2.1 \times 10^{-1}$ & $1.3 \times 10^{-2}$ & $1.3 \times 10^{-2}$ & $4.0 \times 10^{-3}$ & $6.3 \times 10^{-8}$ & $9.1 \times 10^{-13}$ \\
        \midrule
        \multicolumn{8}{c}{Mixed Abs. Energy Errors} \\
        \midrule
        Molnupiravir & $1.9 \times 10^{-4}$ & $6.2 \times 10^{-5}$ & $1.2 \times 10^{-9}$ & $9.8 \times 10^{-10}$ & $2.0 \times 10^{-10}$ & $1.4 \times 10^{-13}$ & $1.1 \times 10^{-13}$ \\
        Water 237 & $1.6 \times 10^{-3}$ & $5.3 \times 10^{-4}$ & $2.3 \times 10^{-8}$ & $1.9 \times 10^{-8}$ & $6.9 \times 10^{-9}$ & $8.0 \times 10^{-13}$ & $1.7 \times 10^{-12}$ \\
        Selenite/A & $8.7 \times 10^{-3}$ & $2.4 \times 10^{-3}$ & $8.4 \times 10^{-8}$ & $8.3 \times 10^{-8}$ & $2.9 \times 10^{-8}$ & $1.6 \times 10^{-11}$ & $0$ \\
        Selenite/B & $8.8 \times 10^{-3}$ & $3.0 \times 10^{-3}$ & $2.4 \times 10^{-7}$ & $2.8 \times 10^{-7}$ & $6.9 \times 10^{-8}$ & $2.3 \times 10^{-11}$ & $4.5 \times 10^{-13}$ \\
        Selenite/C & $5.4 \times 10^{-2}$ & $1.6 \times 10^{-2}$ & $4.2 \times 10^{-6}$ & $4.6 \times 10^{-6}$ & $5.3 \times 10^{-7}$ & $7.5 \times 10^{-11}$ & $9.1 \times 10^{-13}$ \\
        \midrule
        \multicolumn{8}{c}{Purification Iterations} \\
        \midrule
        Molnupiravir & $17$ & $27$ & $27$ & $24$ & $27$ & $13$ & $13$ \\
        Water 237 & $20$ & $45$ & $17$ & $15$ & $13$ & $11$ & $11$ \\
        Selenite/A & $101$ & $74$ & $22$ & $29$ & $23$ & $22$ & $22$ \\
        Selenite/B & $72$ & $101$ & $37$ & $33$ & $30$ & $27$ & $27$ \\
        Selenite/C & $101$ & $66$ & $31$ & $41$ & $32$ & $32$ & $28$ \\
        \bottomrule
        \bottomrule
    \end{tabular}
    \caption{\hl{Errors and number of purification iterations for the various systems computed with FP16 Tensor Cores vs. the numbers of correction terms. Non-converged systems are labeled as NC (not converged) in the number of iterations. A: PCSEG-1; B: AUG-PCSEG-1; C: PCSEG-3. 1 Term: $A^0B^0$; 2 Terms: $A^0B^0 + A^0B^1$; 3 Terms: $A^0B^0 + A^0B^1 + A^1B^0$; 4 Terms: $A^0B^0 + A^0B^1 + A^1B^0 + A^0B^0$. 4-MPFR: the 4 Term MPFR implementation with a double precision exponent. Ozaki (5): Ozaki scheme with 5 splits. FP64: calculations performed with double precision.}}
    \label{tab:markidis}
\end{table*}

While comparing our CUDA and MPFR implementations of the Markidis method, we noted that a significant amount of error was introduced due to the limited exponent range. To improve this, we applied the same scaling method for the exponent ranges used in the Ozaki scheme~\cite{10.1007/978-3-030-50743-5_12}, and found it provided substantial improvement. For example, the 4 Terms result for Selenite/PCSEG-3 improved from an \hl{RMSD} error of $1.72\times10^{-5}$ to $2.86\times10^{-6}$. Nonetheless, the MPFR result \hl{with a double precision exponent} still remains more precise; in the future, it may be possible to further improve the Markidis method with a new scaling scheme.

In the implementation of Finkelstein et al., they include three terms in the Markidis correction, which is well justified from our data. This three term result can be accomplished at the cost of only two multiplications because they implement a scheme that only requires squaring symmetric matrices, which means that one can exploit the relation $A^1A^0 = (A^0A^1)^T$ even in inexact arithmetic. The benefit of the Markidis method is thus that it can substantially improve the accuracy at a low cost with FP16 Tensor Cores (especially if combined with the exponent scaling scheme). On the other hand, it does not converge to the double precision result, can't take advantage of lower precision hardware like FP16 without FP32 accumulation, and has significant errors in more challenging numerical conditions (like larger systems and basis sets). \hl{In particular, the commutator error remains large for the Markidis method, and for the Selenite/C system the energy error is above $10^{-6}$ Hartree even with the iterative refinement approach}. For this reason, the higher order approximation of the Ozaki scheme is valuable. When only the matrix square is required, the Ozaki scheme can also exploit this symmetry in the calculation of the error free terms (see Fig.~\ref{fig:ozaki_over}) as well as other optimizations detailed by Uchino~\cite{uchino_thesis}. \hl{Using this symmetry and only computing the square of the matrix to purify, the Ozaki scheme with four or five terms would require 8 or 11 multiplications respectively (compared to the two required by the Markidis scheme).}

\subsection{Mixed Precision Purification}
\label{sec:comm}

Finally, we will revisit the result of Sec.~\ref{sec:sweep} regarding the precision needed for multiplying matrices and for accumulation. The findings of Fig.~\ref{fig:sweep} indicate that the matrices used in purification may be stored in FP32 as long as the multiplication upcasts them to FP64. This could be utilized in current libraries that implement purification~\cite{Borstnik2014, Mohr2017, Dawson2018, Bock2018, Rubensson2022} to reduce communication and data transfer costs, even without considering the Ozaki scheme. To validate this finding, we modify the NTPoly code~\cite{Dawson2018}, which implements the TRS4 method using sparse matrix algebra. During the multiplication process, we down cast all input matrix elements back and forth from FP32 to simulate the loss of precision. As a test matrix, we use the Hamiltonians coming from a BigDFT calculation of a bulk silicon supercell (3240 atoms) and an NTChem calculation of a water cluster (573 molecules).

\begin{table}[htbp]
    \centering
    \scriptsize
    \begin{tabular}{lccc}
        \toprule
        \toprule
        Precision & \textbf{$\epsilon=1\times 10^{-5}$} & \textbf{$\epsilon=1\times 10^{-6}$} & \textbf{$\epsilon=1\times 10^{-8}$} \\
        \midrule
        \multicolumn{4}{c}{Silicon / BigDFT} \\
        \midrule
        Double Precision & $1.827\times10^{-3}$ & $5.904\times10^{-5}$ & $3.282\times10^{-5}$ \\
        Single Precision & $1.832\times10^{-3}$ & $6.016\times10^{-5}$ & $3.280\times10^{-5}$ \\
        \midrule
        \multicolumn{4}{c}{Water / PCSEG-1} \\
        \midrule
        Double Precision & $1.030\times10^{-3}$ & $3.452\times10^{-5}$ & $2.44\times10^{-7}$ \\
        Single Precision & $1.038\times10^{-3}$ & $3.261\times10^{-5}$ & $5.289\times10^{-5}$ \\
        \bottomrule
        \bottomrule
    \end{tabular}
    \caption{Errors (reference - computed) in energy (Hartree) for different thresholds ($\epsilon$) and precisions when performing density matrix purification with the NTPoly code on bulk silicon from BigDFT and a water cluster from NTChem. }
    \label{tab:thresholds}
\end{table}

We compare for double precision and single precision purification against the electronic energy computed with dense diagonalization in double precision.  Since NTPoly enforces sparsity in the underlying matrices by filtering small values, we report in Tab.~\ref{tab:thresholds} the errors in energy when using several different thresholds. We find that the use of single precision for input matrix values has a negligible effect on the quality of the purification result. For the Silicon / BigDFT calculation, the number of purification iterations remained unchanged; for the Water / PCSEG-1 system, we did observe the iterations increased due to stagnation. We anticipate that the results here may be further improved with a suitable scaling scheme to handle the limited exponent range. These results demonstrate how the findings of Fig.~\ref{fig:sweep} can have an immediate impact of quantum chemistry codes.

\section{Conclusion}

In this work, we have examined the specific floating point precision requirements of a representative kernel from quantum chemistry calculations: calculation of the single particle density matrix using density matrix purification. Exploiting MPFR to emulate arbitrary mantissa sizes, we found that double precision affords an unnecessarily high level of precision. If a precision between single and double precision is used (for example, with an effective 37 bit mantissa), a reliable result can be obtained. We further identified that single precision is sufficient \hl{for many practical problems} if accumulation is done in a higher precision; libraries that implement density matrix purification or similar algorithms may immediately exploit this fact to reduce communication and data transfer costs. 

To further take advantage of this reduced precision requirement, we proposed the use of the Ozaki scheme with a smaller number of splits. We found that the reduced precision requirements of purification leads to a reduction in the number of multiplications needed for the Ozaki scheme by about a factor of two. In this work, we have only examined the reliability of this approach, and not implemented an optimized version. Nonetheless, on an RTX A6000 GPU the Ozaki scheme has already been demonstrated to outperform standard double precision multiplication calls~\cite{ootomo2023dgemm}. Furthermore, on an A100 a similar algorithm to purification was shown to require less time to solution then dense diagonalization~\cite{Finkelstein2023}. Hence, a combination of the two may provide practical benefit in the short term (in particular, a node parallel version). More importantly, if future architectures are developed with an even higher ratio of low precision multiplication to FP64, our work shows that a performance improvement could be realized without sacrificing precision.

The methodology developed here may be straightforwardly applied to a number of other matrix multiplication based algorithms in quantum chemistry --- particularly many-body methods like MP2 and Coupled Cluster. \hl{For materials science applications based on planewaves, finite elements, etc., the Ozaki scheme may be used to accelerate steps like orbital orthogonalization or subspace diagonalization.} Software emulation of low precision results can substantially increase the time to solution, however many practical problems remain in reach, particularly with the development of appropriate libraries for multi-precision linear algebra~\cite{2109.13406v2}. \hl{Another approach may be to introduce artificial numerical noise into codes~\cite{Knizia2011} or} use tools like veritracer to \hl{automatically} instrument electronic structure codes and measure floating point errors~\cite{Chatelain2018, Gavini2023}. 

It will be particularly essential to test low precision algorithms in combination with physically motivated approximations. Skilled practitioners already know how to employ every available approximation (smaller basis sets, lower levels of theory, low-rank approximations, numerical thresholds, spatial locality, etc.) to achieve a reliable result using as little computational resources as possible. Low precision approximations will almost certainly fail to have an improved trade off between cost and accuracy than domain specific methods. Fortunately, the Ozaki scheme represents one low precision approximation that can accelerate calculations without sacrificing meaningful amounts of precision, making it an ideal candidate for combination with the diverse set of algorithms available in quantum chemistry.

\section{Acknowledgements}

Computations were performed using resources at the Research Center for Computational Science, Okazaki, Japan (Projects: 23-IMS-C029 and 24-IMS-C151). We gratefully acknowledge members of the RIKEN R-CCS Low Precision Working Group for their advice and guidance. 

\bibliography{apssamp}

\providecommand{\latin}[1]{#1}
\makeatletter
\providecommand{\doi}
  {\begingroup\let\do\@makeother\dospecials
  \catcode`\{=1 \catcode`\}=2 \doi@aux}
\providecommand{\doi@aux}[1]{\endgroup\texttt{#1}}
\makeatother
\providecommand*\mcitethebibliography{\thebibliography}
\csname @ifundefined\endcsname{endmcitethebibliography}
  {\let\endmcitethebibliography\endthebibliography}{}
\begin{mcitethebibliography}{83}
\providecommand*\natexlab[1]{#1}
\providecommand*\mciteSetBstSublistMode[1]{}
\providecommand*\mciteSetBstMaxWidthForm[2]{}
\providecommand*\mciteBstWouldAddEndPuncttrue
  {\def\EndOfBibitem{\unskip.}}
\providecommand*\mciteBstWouldAddEndPunctfalse
  {\let\EndOfBibitem\relax}
\providecommand*\mciteSetBstMidEndSepPunct[3]{}
\providecommand*\mciteSetBstSublistLabelBeginEnd[3]{}
\providecommand*\EndOfBibitem{}
\mciteSetBstSublistMode{f}
\mciteSetBstMaxWidthForm{subitem}{(\alph{mcitesubitemcount})}
\mciteSetBstSublistLabelBeginEnd
  {\mcitemaxwidthsubitemform\space}
  {\relax}
  {\relax}

\bibitem[Sevilla \latin{et~al.}(2023)Sevilla, Ho, and
  Besiroglu]{sevilla2023please}
Sevilla,~J.; Ho,~A.; Besiroglu,~T. Please Report Your Compute.
  \emph{Communications of the ACM} \textbf{2023}, \emph{66}, 30--32\relax
\mciteBstWouldAddEndPuncttrue
\mciteSetBstMidEndSepPunct{\mcitedefaultmidpunct}
{\mcitedefaultendpunct}{\mcitedefaultseppunct}\relax
\EndOfBibitem
\bibitem[Markidis \latin{et~al.}(2018)Markidis, Der~Chien, Laure, Peng, and
  Vetter]{markidis2018nvidia}
Markidis,~S.; Der~Chien,~S.~W.; Laure,~E.; Peng,~I.~B.; Vetter,~J.~S. Nvidia
  tensor core programmability, performance \& precision. 2018 IEEE
  international parallel and distributed processing symposium workshops
  (IPDPSW). 2018; pp 522--531\relax
\mciteBstWouldAddEndPuncttrue
\mciteSetBstMidEndSepPunct{\mcitedefaultmidpunct}
{\mcitedefaultendpunct}{\mcitedefaultseppunct}\relax
\EndOfBibitem
\bibitem[Lewis \latin{et~al.}(2022)Lewis, Beall, Ganahl, Hauru, Mallick, and
  Vidal]{Lewis2022}
Lewis,~A. G.~M.; Beall,~J.; Ganahl,~M.; Hauru,~M.; Mallick,~S.~B.; Vidal,~G.
  Large-scale distributed linear algebra with tensor processing units.
  \emph{Proceedings of the National Academy of Sciences} \textbf{2022},
  \emph{119}, e2122762119\relax
\mciteBstWouldAddEndPuncttrue
\mciteSetBstMidEndSepPunct{\mcitedefaultmidpunct}
{\mcitedefaultendpunct}{\mcitedefaultseppunct}\relax
\EndOfBibitem
\bibitem[Ozaki \latin{et~al.}(2012)Ozaki, Ogita, Oishi, and
  Rump]{ozaki2012error}
Ozaki,~K.; Ogita,~T.; Oishi,~S.; Rump,~S.~M. Error-free transformations of
  matrix multiplication by using fast routines of matrix multiplication and its
  applications. \emph{Numerical Algorithms} \textbf{2012}, \emph{59},
  95--118\relax
\mciteBstWouldAddEndPuncttrue
\mciteSetBstMidEndSepPunct{\mcitedefaultmidpunct}
{\mcitedefaultendpunct}{\mcitedefaultseppunct}\relax
\EndOfBibitem
\bibitem[Palser and Manolopoulos(1998)Palser, and
  Manolopoulos]{palser1998canonical}
Palser,~A.~H.; Manolopoulos,~D.~E. Canonical purification of the density matrix
  in electronic-structure theory. \emph{Physical Review B} \textbf{1998},
  \emph{58}, 12704\relax
\mciteBstWouldAddEndPuncttrue
\mciteSetBstMidEndSepPunct{\mcitedefaultmidpunct}
{\mcitedefaultendpunct}{\mcitedefaultseppunct}\relax
\EndOfBibitem
\bibitem[{NVIDIA Corporation}(2018)]{nvidia_v100_datasheet}
{NVIDIA Corporation} NVIDIA Tesla V100 GPU Architecture: The World’s Most
  Advanced Data Center GPU.
  \url{https://images.nvidia.com/content/technologies/volta/pdf/tesla-volta-v100-datasheet-letter-fnl-web.pdf},
  2018; Accessed: 2024-09-03\relax
\mciteBstWouldAddEndPuncttrue
\mciteSetBstMidEndSepPunct{\mcitedefaultmidpunct}
{\mcitedefaultendpunct}{\mcitedefaultseppunct}\relax
\EndOfBibitem
\bibitem[{NVIDIA Corporation}(2021)]{nvidia_a100_datasheet}
{NVIDIA Corporation} NVIDIA A100 Tensor Core GPU Architecture: Unprecedented
  Acceleration at Every Scale.
  \url{https://www.nvidia.com/content/dam/en-zz/Solutions/Data-Center/a100/pdf/nvidia-a100-datasheet-us-nvidia-1758950-r4-web.pdf},
  2021; Accessed: 2024-09-03\relax
\mciteBstWouldAddEndPuncttrue
\mciteSetBstMidEndSepPunct{\mcitedefaultmidpunct}
{\mcitedefaultendpunct}{\mcitedefaultseppunct}\relax
\EndOfBibitem
\bibitem[{NVIDIA Corporation}(2024)]{nvidia_h100_datasheet}
{NVIDIA Corporation} NVIDIA H100 Tensor Core GPU: Extraordinary performance,
  scalability, and security for every data center.
  \url{https://resources.nvidia.com/en-us-tensor-core/nvidia-tensor-core-gpu-datasheet},
  2024; Accessed: 2024-09-03\relax
\mciteBstWouldAddEndPuncttrue
\mciteSetBstMidEndSepPunct{\mcitedefaultmidpunct}
{\mcitedefaultendpunct}{\mcitedefaultseppunct}\relax
\EndOfBibitem
\bibitem[{HPC Systems Inc.}(2022)]{gpu_list}
{HPC Systems Inc.} GPU Performance Specifications Comparison Table.
  \url{https://www.hpc.co.jp/product/wp-content/uploads/sites/3/2022/07/GPU-list_A3.pdf},
  2022; Accessed: 2024-09-06\relax
\mciteBstWouldAddEndPuncttrue
\mciteSetBstMidEndSepPunct{\mcitedefaultmidpunct}
{\mcitedefaultendpunct}{\mcitedefaultseppunct}\relax
\EndOfBibitem
\bibitem[Yasuda(2008)]{Yasuda2008}
Yasuda,~K. Two-electron integral evaluation on the graphics processor unit.
  \emph{Journal of Computational Chemistry} \textbf{2008}, \emph{29},
  334--342\relax
\mciteBstWouldAddEndPuncttrue
\mciteSetBstMidEndSepPunct{\mcitedefaultmidpunct}
{\mcitedefaultendpunct}{\mcitedefaultseppunct}\relax
\EndOfBibitem
\bibitem[Ufimtsev and Mart{\'i}nez(2008)Ufimtsev, and
  Mart{\'i}nez]{Ufimtsev2008}
Ufimtsev,~I.~S.; Mart{\'i}nez,~T.~J. Quantum Chemistry on Graphical Processing
  Units. 1. Strategies for Two-Electron Integral Evaluation. \emph{Journal of
  Chemical Theory and Computation} \textbf{2008}, \emph{4}, 222--231\relax
\mciteBstWouldAddEndPuncttrue
\mciteSetBstMidEndSepPunct{\mcitedefaultmidpunct}
{\mcitedefaultendpunct}{\mcitedefaultseppunct}\relax
\EndOfBibitem
\bibitem[Luehr \latin{et~al.}(2011)Luehr, Ufimtsev, and
  Mart{\'i}nez]{Luehr2011}
Luehr,~N.; Ufimtsev,~I.~S.; Mart{\'i}nez,~T.~J. Dynamic Precision for Electron
  Repulsion Integral Evaluation on Graphical Processing Units (GPUs).
  \emph{Journal of Chemical Theory and Computation} \textbf{2011}, \emph{7},
  949--954\relax
\mciteBstWouldAddEndPuncttrue
\mciteSetBstMidEndSepPunct{\mcitedefaultmidpunct}
{\mcitedefaultendpunct}{\mcitedefaultseppunct}\relax
\EndOfBibitem
\bibitem[Tornai \latin{et~al.}(2019)Tornai, Ladj{\'a}nszki, R{\'a}k, Kis, and
  Cserey]{Tornai2019}
Tornai,~G.~J.; Ladj{\'a}nszki,~I.; R{\'a}k,~{\'A}.; Kis,~G.; Cserey,~G.
  Calculation of Quantum Chemical Two-Electron Integrals by Applying Compiler
  Technology on GPU. \emph{Journal of Chemical Theory and Computation}
  \textbf{2019}, \emph{15}, 5319--5331\relax
\mciteBstWouldAddEndPuncttrue
\mciteSetBstMidEndSepPunct{\mcitedefaultmidpunct}
{\mcitedefaultendpunct}{\mcitedefaultseppunct}\relax
\EndOfBibitem
\bibitem[Laqua \latin{et~al.}(2021)Laqua, Kussmann, and Ochsenfeld]{Laqua2021}
Laqua,~H.; Kussmann,~J.; Ochsenfeld,~C. Accelerating seminumerical
  Fock-exchange calculations using mixed single- and double-precision
  arithmethic. \emph{The Journal of Chemical Physics} \textbf{2021},
  \emph{154}, 214116\relax
\mciteBstWouldAddEndPuncttrue
\mciteSetBstMidEndSepPunct{\mcitedefaultmidpunct}
{\mcitedefaultendpunct}{\mcitedefaultseppunct}\relax
\EndOfBibitem
\bibitem[Dang \latin{et~al.}(2022)Dang, Wilson, and
  Zimmerman]{dang2022numerical}
Dang,~D.-K.; Wilson,~L.~W.; Zimmerman,~P.~M. The numerical evaluation of Slater
  integrals on graphics processing units. \emph{Journal of Computational
  Chemistry} \textbf{2022}, \emph{43}, 1680--1689\relax
\mciteBstWouldAddEndPuncttrue
\mciteSetBstMidEndSepPunct{\mcitedefaultmidpunct}
{\mcitedefaultendpunct}{\mcitedefaultseppunct}\relax
\EndOfBibitem
\bibitem[Vinson(2020)]{Vinson2020}
Vinson,~J. Faster exact exchange in periodic systems using single-precision
  arithmetic. \emph{The Journal of Chemical Physics} \textbf{2020}, \emph{153},
  204106\relax
\mciteBstWouldAddEndPuncttrue
\mciteSetBstMidEndSepPunct{\mcitedefaultmidpunct}
{\mcitedefaultendpunct}{\mcitedefaultseppunct}\relax
\EndOfBibitem
\bibitem[Tsuchida and Choe(2012)Tsuchida, and Choe]{Tsuchida2012}
Tsuchida,~E.; Choe,~Y.-K. Iterative diagonalization of symmetric matrices in
  mixed precision and its application to electronic structure calculations.
  \emph{Computer Physics Communications} \textbf{2012}, \emph{183},
  980--985\relax
\mciteBstWouldAddEndPuncttrue
\mciteSetBstMidEndSepPunct{\mcitedefaultmidpunct}
{\mcitedefaultendpunct}{\mcitedefaultseppunct}\relax
\EndOfBibitem
\bibitem[Das \latin{et~al.}(2019)Das, Motamarri, Gavini, Turcksin, Li, and
  Leback]{10.1145/3295500.3357157}
Das,~S.; Motamarri,~P.; Gavini,~V.; Turcksin,~B.; Li,~Y.~W.; Leback,~B. Fast,
  Scalable and Accurate Finite-Element Based Ab Initio Calculations Using Mixed
  Precision Computing: 46 PFLOPS Simulation of a Metallic Dislocation System.
  Proceedings of the International Conference for High Performance Computing,
  Networking, Storage and Analysis. New York, NY, USA, 2019\relax
\mciteBstWouldAddEndPuncttrue
\mciteSetBstMidEndSepPunct{\mcitedefaultmidpunct}
{\mcitedefaultendpunct}{\mcitedefaultseppunct}\relax
\EndOfBibitem
\bibitem[Das \latin{et~al.}(2022)Das, Motamarri, Subramanian, Rogers, and
  Gavini]{Das2022}
Das,~S.; Motamarri,~P.; Subramanian,~V.; Rogers,~D.~M.; Gavini,~V. DFT-FE 1.0:
  A massively parallel hybrid CPU-GPU density functional theory code using
  finite-element discretization. \emph{Computer Physics Communications}
  \textbf{2022}, \emph{280}, 108473\relax
\mciteBstWouldAddEndPuncttrue
\mciteSetBstMidEndSepPunct{\mcitedefaultmidpunct}
{\mcitedefaultendpunct}{\mcitedefaultseppunct}\relax
\EndOfBibitem
\bibitem[Das \latin{et~al.}(2023)Das, Kanungo, Subramanian, Panigrahi,
  Motamarri, Rogers, Zimmerman, and Gavini]{10.1145/3581784.3627037}
Das,~S.; Kanungo,~B.; Subramanian,~V.; Panigrahi,~G.; Motamarri,~P.;
  Rogers,~D.; Zimmerman,~P.; Gavini,~V. Large-Scale Materials Modeling at
  Quantum Accuracy: Ab Initio Simulations of Quasicrystals and Interacting
  Extended Defects in Metallic Alloys. Proceedings of the International
  Conference for High Performance Computing, Networking, Storage and Analysis.
  New York, NY, USA, 2023\relax
\mciteBstWouldAddEndPuncttrue
\mciteSetBstMidEndSepPunct{\mcitedefaultmidpunct}
{\mcitedefaultendpunct}{\mcitedefaultseppunct}\relax
\EndOfBibitem
\bibitem[Woo \latin{et~al.}(2023)Woo, Kim, and Kim]{Woo2023}
Woo,~J.; Kim,~S.; Kim,~W.~Y. Dynamic Precision Approach for Accelerating
  Large-Scale Eigenvalue Solvers in Electronic Structure Calculations on
  Graphics Processing Units. \emph{Journal of Chemical Theory and Computation}
  \textbf{2023}, \emph{19}, 1457--1465\relax
\mciteBstWouldAddEndPuncttrue
\mciteSetBstMidEndSepPunct{\mcitedefaultmidpunct}
{\mcitedefaultendpunct}{\mcitedefaultseppunct}\relax
\EndOfBibitem
\bibitem[Khadatkar and Motamarri(2023)Khadatkar, and Motamarri]{Khadatkar2023}
Khadatkar,~S.; Motamarri,~P. Subspace recursive Fermi-operator expansion
  strategies for large-scale DFT eigenvalue problems on HPC architectures.
  \emph{The Journal of Chemical Physics} \textbf{2023}, \emph{159},
  031102\relax
\mciteBstWouldAddEndPuncttrue
\mciteSetBstMidEndSepPunct{\mcitedefaultmidpunct}
{\mcitedefaultendpunct}{\mcitedefaultseppunct}\relax
\EndOfBibitem
\bibitem[Vogt \latin{et~al.}(2008)Vogt, Olivares-Amaya, Kermes, Shao,
  Amador-Bedolla, and Aspuru-Guzik]{Vogt2008}
Vogt,~L.; Olivares-Amaya,~R.; Kermes,~S.; Shao,~Y.; Amador-Bedolla,~C.;
  Aspuru-Guzik,~A. Accelerating Resolution-of-the-Identity Second-Order
  M{\o}ller-Plesset Quantum Chemistry Calculations with Graphical Processing
  Units. \emph{The Journal of Physical Chemistry A} \textbf{2008}, \emph{112},
  2049--2057\relax
\mciteBstWouldAddEndPuncttrue
\mciteSetBstMidEndSepPunct{\mcitedefaultmidpunct}
{\mcitedefaultendpunct}{\mcitedefaultseppunct}\relax
\EndOfBibitem
\bibitem[Olivares-Amaya \latin{et~al.}(2010)Olivares-Amaya, Watson, Edgar,
  Vogt, Shao, and Aspuru-Guzik]{Olivares-Amaya2010}
Olivares-Amaya,~R.; Watson,~M.~A.; Edgar,~R.~G.; Vogt,~L.; Shao,~Y.;
  Aspuru-Guzik,~A. Accelerating Correlated Quantum Chemistry Calculations Using
  Graphical Processing Units and a Mixed Precision Matrix Multiplication
  Library. \emph{Journal of Chemical Theory and Computation} \textbf{2010},
  \emph{6}, 135--144\relax
\mciteBstWouldAddEndPuncttrue
\mciteSetBstMidEndSepPunct{\mcitedefaultmidpunct}
{\mcitedefaultendpunct}{\mcitedefaultseppunct}\relax
\EndOfBibitem
\bibitem[Vysotskiy and Cederbaum(2011)Vysotskiy, and Cederbaum]{Vysotskiy2011}
Vysotskiy,~V.~P.; Cederbaum,~L.~S. Accurate Quantum Chemistry in Single
  Precision Arithmetic: Correlation Energy. \emph{Journal of Chemical Theory
  and Computation} \textbf{2011}, \emph{7}, 320--326\relax
\mciteBstWouldAddEndPuncttrue
\mciteSetBstMidEndSepPunct{\mcitedefaultmidpunct}
{\mcitedefaultendpunct}{\mcitedefaultseppunct}\relax
\EndOfBibitem
\bibitem[DePrince~III and Hammond(2011)DePrince~III, and Hammond]{DePrince2011}
DePrince~III,~A.~E.; Hammond,~J.~R. Coupled Cluster Theory on Graphics
  Processing Units I. The Coupled Cluster Doubles Method. \emph{Journal of
  Chemical Theory and Computation} \textbf{2011}, \emph{7}, 1287--1295\relax
\mciteBstWouldAddEndPuncttrue
\mciteSetBstMidEndSepPunct{\mcitedefaultmidpunct}
{\mcitedefaultendpunct}{\mcitedefaultseppunct}\relax
\EndOfBibitem
\bibitem[Pokhilko \latin{et~al.}(2018)Pokhilko, Epifanovsky, and
  Krylov]{Pokhilko2018}
Pokhilko,~P.; Epifanovsky,~E.; Krylov,~A.~I. Double Precision Is Not Needed for
  Many-Body Calculations: Emergent Conventional Wisdom. \emph{Journal of
  Chemical Theory and Computation} \textbf{2018}, \emph{14}, 4088--4096\relax
\mciteBstWouldAddEndPuncttrue
\mciteSetBstMidEndSepPunct{\mcitedefaultmidpunct}
{\mcitedefaultendpunct}{\mcitedefaultseppunct}\relax
\EndOfBibitem
\bibitem[Wang \latin{et~al.}(2022)Wang, Peyton, and Crawford]{Wang2022}
Wang,~Z.; Peyton,~B.~G.; Crawford,~T.~D. Accelerating Real-Time Coupled Cluster
  Methods with Single-Precision Arithmetic and Adaptive Numerical Integration.
  \emph{Journal of Chemical Theory and Computation} \textbf{2022}, \emph{18},
  5479--5491\relax
\mciteBstWouldAddEndPuncttrue
\mciteSetBstMidEndSepPunct{\mcitedefaultmidpunct}
{\mcitedefaultendpunct}{\mcitedefaultseppunct}\relax
\EndOfBibitem
\bibitem[Tian \latin{et~al.}(2022)Tian, Xie, Luo, and Ma]{Tian2022}
Tian,~Y.; Xie,~Z.; Luo,~Z.; Ma,~H. Mixed-Precision Implementation of the
  Density Matrix Renormalization Group. \emph{Journal of Chemical Theory and
  Computation} \textbf{2022}, \emph{18}, 7260--7271\relax
\mciteBstWouldAddEndPuncttrue
\mciteSetBstMidEndSepPunct{\mcitedefaultmidpunct}
{\mcitedefaultendpunct}{\mcitedefaultseppunct}\relax
\EndOfBibitem
\bibitem[Ziogas \latin{et~al.}(2019)Ziogas, Ben-Nun, Fern\'{a}ndez, Schneider,
  Luisier, and Hoefler]{10.1145/3295500.3357156}
Ziogas,~A.~N.; Ben-Nun,~T.; Fern\'{a}ndez,~G.~I.; Schneider,~T.; Luisier,~M.;
  Hoefler,~T. A Data-Centric Approach to Extreme-Scale Ab Initio Dissipative
  Quantum Transport Simulations. Proceedings of the International Conference
  for High Performance Computing, Networking, Storage and Analysis. New York,
  NY, USA, 2019\relax
\mciteBstWouldAddEndPuncttrue
\mciteSetBstMidEndSepPunct{\mcitedefaultmidpunct}
{\mcitedefaultendpunct}{\mcitedefaultseppunct}\relax
\EndOfBibitem
\bibitem[Yu and Govoni(2022)Yu, and Govoni]{Yu2022}
Yu,~V. W.-z.; Govoni,~M. GPU Acceleration of Large-Scale Full-Frequency GW
  Calculations. \emph{Journal of Chemical Theory and Computation}
  \textbf{2022}, \emph{18}, 4690--4707\relax
\mciteBstWouldAddEndPuncttrue
\mciteSetBstMidEndSepPunct{\mcitedefaultmidpunct}
{\mcitedefaultendpunct}{\mcitedefaultseppunct}\relax
\EndOfBibitem
\bibitem[Yasuda(2008)]{Yasuda20082}
Yasuda,~K. Accelerating Density Functional Calculations with Graphics
  Processing Unit. \emph{Journal of Chemical Theory and Computation}
  \textbf{2008}, \emph{4}, 1230--1236\relax
\mciteBstWouldAddEndPuncttrue
\mciteSetBstMidEndSepPunct{\mcitedefaultmidpunct}
{\mcitedefaultendpunct}{\mcitedefaultseppunct}\relax
\EndOfBibitem
\bibitem[Williams-Young \latin{et~al.}(2020)Williams-Young, de~Jong, van Dam,
  and Yang]{10.3389/fchem.2020.581058}
Williams-Young,~D.~B.; de~Jong,~W.~A.; van Dam,~H. J.~J.; Yang,~C. On the
  Efficient Evaluation of the Exchange Correlation Potential on Graphics
  Processing Unit Clusters. \emph{Frontiers in Chemistry} \textbf{2020},
  \emph{8}\relax
\mciteBstWouldAddEndPuncttrue
\mciteSetBstMidEndSepPunct{\mcitedefaultmidpunct}
{\mcitedefaultendpunct}{\mcitedefaultseppunct}\relax
\EndOfBibitem
\bibitem[Hohenberg and Kohn(1964)Hohenberg, and
  Kohn]{hohenberg-inhomogeneous-1964}
Hohenberg,~P.; Kohn,~W. Inhomogeneous Electron Gas. \emph{Phys. Rev.}
  \textbf{1964}, \emph{136}, B864--B871\relax
\mciteBstWouldAddEndPuncttrue
\mciteSetBstMidEndSepPunct{\mcitedefaultmidpunct}
{\mcitedefaultendpunct}{\mcitedefaultseppunct}\relax
\EndOfBibitem
\bibitem[Kohn and Sham(1965)Kohn, and Sham]{kohn-self_consistent-1965}
Kohn,~W.; Sham,~L.~J. Self-Consistent Equations Including Exchange and
  Correlation Effects. \emph{Phys. Rev.} \textbf{1965}, \emph{140},
  A1133--A1138\relax
\mciteBstWouldAddEndPuncttrue
\mciteSetBstMidEndSepPunct{\mcitedefaultmidpunct}
{\mcitedefaultendpunct}{\mcitedefaultseppunct}\relax
\EndOfBibitem
\bibitem[Bowler and Miyazaki(2012)Bowler, and Miyazaki]{bowler-ON-2012}
Bowler,~D.~R.; Miyazaki,~T. {O(N) methods in electronic structure
  calculations}. \emph{Rep. Prog. Phys.} \textbf{2012}, \emph{75}, 036503\relax
\mciteBstWouldAddEndPuncttrue
\mciteSetBstMidEndSepPunct{\mcitedefaultmidpunct}
{\mcitedefaultendpunct}{\mcitedefaultseppunct}\relax
\EndOfBibitem
\bibitem[McWeeny(1960)]{RevModPhys.32.335}
McWeeny,~R. Some Recent Advances in Density Matrix Theory. \emph{Rev. Mod.
  Phys.} \textbf{1960}, \emph{32}, 335--369\relax
\mciteBstWouldAddEndPuncttrue
\mciteSetBstMidEndSepPunct{\mcitedefaultmidpunct}
{\mcitedefaultendpunct}{\mcitedefaultseppunct}\relax
\EndOfBibitem
\bibitem[Prodan and Kohn(2005)Prodan, and Kohn]{Prodan2005}
Prodan,~E.; Kohn,~W. Nearsightedness of electronic matter. \emph{Proceedings of
  the National Academy of Sciences} \textbf{2005}, \emph{102},
  11635--11638\relax
\mciteBstWouldAddEndPuncttrue
\mciteSetBstMidEndSepPunct{\mcitedefaultmidpunct}
{\mcitedefaultendpunct}{\mcitedefaultseppunct}\relax
\EndOfBibitem
\bibitem[Dawson and Nakajima(2018)Dawson, and Nakajima]{Dawson2018}
Dawson,~W.; Nakajima,~T. Massively parallel sparse matrix function calculations
  with NTPoly. \emph{Computer Physics Communications} \textbf{2018},
  \emph{225}, 154--165\relax
\mciteBstWouldAddEndPuncttrue
\mciteSetBstMidEndSepPunct{\mcitedefaultmidpunct}
{\mcitedefaultendpunct}{\mcitedefaultseppunct}\relax
\EndOfBibitem
\bibitem[Niklasson \latin{et~al.}(2003)Niklasson, Tymczak, and
  Challacombe]{Niklasson2003}
Niklasson,~A. M.~N.; Tymczak,~C.~J.; Challacombe,~M. Trace resetting density
  matrix purification in O(N) self-consistent-field theory. \emph{The Journal
  of Chemical Physics} \textbf{2003}, \emph{118}, 8611--8620\relax
\mciteBstWouldAddEndPuncttrue
\mciteSetBstMidEndSepPunct{\mcitedefaultmidpunct}
{\mcitedefaultendpunct}{\mcitedefaultseppunct}\relax
\EndOfBibitem
\bibitem[Chow \latin{et~al.}(2015)Chow, Liu, Smelyanskiy, and
  Hammond]{Chow2015}
Chow,~E.; Liu,~X.; Smelyanskiy,~M.; Hammond,~J.~R. Parallel scalability of
  Hartree--Fock calculations. \emph{The Journal of Chemical Physics}
  \textbf{2015}, \emph{142}, 104103\relax
\mciteBstWouldAddEndPuncttrue
\mciteSetBstMidEndSepPunct{\mcitedefaultmidpunct}
{\mcitedefaultendpunct}{\mcitedefaultseppunct}\relax
\EndOfBibitem
\bibitem[Chow \latin{et~al.}(2016)Chow, Liu, Misra, Dukhan, Smelyanskiy,
  Hammond, Du, Liao, and Dubey]{Chow2016}
Chow,~E.; Liu,~X.; Misra,~S.; Dukhan,~M.; Smelyanskiy,~M.; Hammond,~J.~R.;
  Du,~Y.; Liao,~X.-K.; Dubey,~P. Scaling up Hartree--Fock calculations on
  Tianhe-2. \emph{The International Journal of High Performance Computing
  Applications} \textbf{2016}, \emph{30}, 85--102\relax
\mciteBstWouldAddEndPuncttrue
\mciteSetBstMidEndSepPunct{\mcitedefaultmidpunct}
{\mcitedefaultendpunct}{\mcitedefaultseppunct}\relax
\EndOfBibitem
\bibitem[Finkelstein \latin{et~al.}(2023)Finkelstein, Negre, and
  Fattebert]{Finkelstein2023}
Finkelstein,~J.; Negre,~C. F.~A.; Fattebert,~J.-L. A fast, dense Chebyshev
  solver for electronic structure on GPUs. \emph{The Journal of Chemical
  Physics} \textbf{2023}, \emph{159}, 101101\relax
\mciteBstWouldAddEndPuncttrue
\mciteSetBstMidEndSepPunct{\mcitedefaultmidpunct}
{\mcitedefaultendpunct}{\mcitedefaultseppunct}\relax
\EndOfBibitem
\bibitem[Pederson \latin{et~al.}(2023)Pederson, Kozlowski, Song, Beall, Ganahl,
  Hauru, Lewis, Yao, Mallick, Blum, and Vidal]{Pederson2023}
Pederson,~R.; Kozlowski,~J.; Song,~R.; Beall,~J.; Ganahl,~M.; Hauru,~M.;
  Lewis,~A. G.~M.; Yao,~Y.; Mallick,~S.~B.; Blum,~V.; Vidal,~G. Large Scale
  Quantum Chemistry with Tensor Processing Units. \emph{Journal of Chemical
  Theory and Computation} \textbf{2023}, \emph{19}, 25--32\relax
\mciteBstWouldAddEndPuncttrue
\mciteSetBstMidEndSepPunct{\mcitedefaultmidpunct}
{\mcitedefaultendpunct}{\mcitedefaultseppunct}\relax
\EndOfBibitem
\bibitem[Finkelstein \latin{et~al.}(2021)Finkelstein, Smith, Mniszewski,
  Barros, Negre, Rubensson, and Niklasson]{Finkelstein2021}
Finkelstein,~J.; Smith,~J.~S.; Mniszewski,~S.~M.; Barros,~K.; Negre,~C. F.~A.;
  Rubensson,~E.~H.; Niklasson,~A. M.~N. Mixed Precision Fermi-Operator
  Expansion on Tensor Cores from a Machine Learning Perspective. \emph{Journal
  of Chemical Theory and Computation} \textbf{2021}, \emph{17},
  2256--2265\relax
\mciteBstWouldAddEndPuncttrue
\mciteSetBstMidEndSepPunct{\mcitedefaultmidpunct}
{\mcitedefaultendpunct}{\mcitedefaultseppunct}\relax
\EndOfBibitem
\bibitem[Finkelstein \latin{et~al.}(2021)Finkelstein, Smith, Mniszewski,
  Barros, Negre, Rubensson, and Niklasson]{Finkelstein20212}
Finkelstein,~J.; Smith,~J.~S.; Mniszewski,~S.~M.; Barros,~K.; Negre,~C. F.~A.;
  Rubensson,~E.~H.; Niklasson,~A. M.~N. Quantum-Based Molecular Dynamics
  Simulations Using Tensor Cores. \emph{Journal of Chemical Theory and
  Computation} \textbf{2021}, \emph{17}, 6180--6192\relax
\mciteBstWouldAddEndPuncttrue
\mciteSetBstMidEndSepPunct{\mcitedefaultmidpunct}
{\mcitedefaultendpunct}{\mcitedefaultseppunct}\relax
\EndOfBibitem
\bibitem[Finkelstein \latin{et~al.}(2022)Finkelstein, Rubensson, Mniszewski,
  Negre, and Niklasson]{Finkelstein2022}
Finkelstein,~J.; Rubensson,~E.~H.; Mniszewski,~S.~M.; Negre,~C. F.~A.;
  Niklasson,~A. M.~N. Quantum Perturbation Theory Using Tensor Cores and a Deep
  Neural Network. \emph{Journal of Chemical Theory and Computation}
  \textbf{2022}, \emph{18}, 4255--4268\relax
\mciteBstWouldAddEndPuncttrue
\mciteSetBstMidEndSepPunct{\mcitedefaultmidpunct}
{\mcitedefaultendpunct}{\mcitedefaultseppunct}\relax
\EndOfBibitem
\bibitem[Habib \latin{et~al.}(2024)Habib, Finkelstein, and
  Niklasson]{Habib2024}
Habib,~A.; Finkelstein,~J.; Niklasson,~A. M.~N. Efficient Mixed-Precision
  Matrix Factorization of the Inverse Overlap Matrix in Electronic Structure
  Calculations with AI-Hardware and GPUs. \emph{Journal of Chemical Theory and
  Computation} \textbf{2024}, \emph{20}, 7102--7112\relax
\mciteBstWouldAddEndPuncttrue
\mciteSetBstMidEndSepPunct{\mcitedefaultmidpunct}
{\mcitedefaultendpunct}{\mcitedefaultseppunct}\relax
\EndOfBibitem
\bibitem[Aarons \latin{et~al.}(2016)Aarons, Sarwar, Thompsett, and
  Skylaris]{Aarons2016}
Aarons,~J.; Sarwar,~M.; Thompsett,~D.; Skylaris,~C.-K. Perspective: Methods for
  large-scale density functional calculations on metallic systems. \emph{The
  Journal of Chemical Physics} \textbf{2016}, \emph{145}, 220901\relax
\mciteBstWouldAddEndPuncttrue
\mciteSetBstMidEndSepPunct{\mcitedefaultmidpunct}
{\mcitedefaultendpunct}{\mcitedefaultseppunct}\relax
\EndOfBibitem
\bibitem[Aarons and Skylaris(2018)Aarons, and Skylaris]{Aarons2018}
Aarons,~J.; Skylaris,~C.-K. Electronic annealing Fermi operator expansion for
  DFT calculations on metallic systems. \emph{The Journal of Chemical Physics}
  \textbf{2018}, \emph{148}, 074107\relax
\mciteBstWouldAddEndPuncttrue
\mciteSetBstMidEndSepPunct{\mcitedefaultmidpunct}
{\mcitedefaultendpunct}{\mcitedefaultseppunct}\relax
\EndOfBibitem
\bibitem[Mniszewski \latin{et~al.}(2019)Mniszewski, Perriot, Rubensson, Negre,
  Cawkwell, and Niklasson]{Mniszewski2019}
Mniszewski,~S.~M.; Perriot,~R.; Rubensson,~E.~H.; Negre,~C. F.~A.;
  Cawkwell,~M.~J.; Niklasson,~A. M.~N. Linear Scaling Pseudo Fermi-Operator
  Expansion for Fractional Occupation. \emph{Journal of Chemical Theory and
  Computation} \textbf{2019}, \emph{15}, 190--200\relax
\mciteBstWouldAddEndPuncttrue
\mciteSetBstMidEndSepPunct{\mcitedefaultmidpunct}
{\mcitedefaultendpunct}{\mcitedefaultseppunct}\relax
\EndOfBibitem
\bibitem[Leamer \latin{et~al.}(2024)Leamer, Dawson, and Bondar]{Leamer2024}
Leamer,~J.~M.; Dawson,~W.; Bondar,~D.~I. Positivity preserving density matrix
  minimization at finite temperatures via square root. \emph{The Journal of
  Chemical Physics} \textbf{2024}, \emph{160}, 074107\relax
\mciteBstWouldAddEndPuncttrue
\mciteSetBstMidEndSepPunct{\mcitedefaultmidpunct}
{\mcitedefaultendpunct}{\mcitedefaultseppunct}\relax
\EndOfBibitem
\bibitem[Goedecker and Teter(1995)Goedecker, and
  Teter]{10.1103/PhysRevB.51.9455}
Goedecker,~S.; Teter,~M. Tight-binding electronic-structure calculations and
  tight-binding molecular dynamics with localized orbitals. \emph{Physical
  Review B} \textbf{1995}, \emph{51}, 9455--9464\relax
\mciteBstWouldAddEndPuncttrue
\mciteSetBstMidEndSepPunct{\mcitedefaultmidpunct}
{\mcitedefaultendpunct}{\mcitedefaultseppunct}\relax
\EndOfBibitem
\bibitem[Mukunoki \latin{et~al.}(2020)Mukunoki, Ogita, and
  Ozaki]{10.1007/978-3-030-43229-4_44}
Mukunoki,~D.; Ogita,~T.; Ozaki,~K. Reproducible BLAS Routines with Tunable
  Accuracy Using Ozaki Scheme for Many-Core Architectures. Parallel Processing
  and Applied Mathematics. Cham, 2020; pp 516--527\relax
\mciteBstWouldAddEndPuncttrue
\mciteSetBstMidEndSepPunct{\mcitedefaultmidpunct}
{\mcitedefaultendpunct}{\mcitedefaultseppunct}\relax
\EndOfBibitem
\bibitem[Ichimura \latin{et~al.}(2018)Ichimura, Katagiri, Ozaki, Ogita, and
  Nagai]{Ichimura2018}
Ichimura,~S.; Katagiri,~T.; Ozaki,~K.; Ogita,~T.; Nagai,~T. Threaded Accurate
  Matrix-Matrix Multiplications with Sparse Matrix-Vector Multiplications. 2018
  IEEE International Parallel and Distributed Processing Symposium Workshops
  (IPDPSW). 2018; pp 1093--1102\relax
\mciteBstWouldAddEndPuncttrue
\mciteSetBstMidEndSepPunct{\mcitedefaultmidpunct}
{\mcitedefaultendpunct}{\mcitedefaultseppunct}\relax
\EndOfBibitem
\bibitem[Mukunoki \latin{et~al.}(2020)Mukunoki, Ozaki, Ogita, and
  Imamura]{10.1007/978-3-030-50743-5_12}
Mukunoki,~D.; Ozaki,~K.; Ogita,~T.; Imamura,~T. DGEMM Using Tensor Cores, and
  Its Accurate and Reproducible Versions. High Performance Computing. Cham,
  2020; pp 230--248\relax
\mciteBstWouldAddEndPuncttrue
\mciteSetBstMidEndSepPunct{\mcitedefaultmidpunct}
{\mcitedefaultendpunct}{\mcitedefaultseppunct}\relax
\EndOfBibitem
\bibitem[Ootomo \latin{et~al.}(2024)Ootomo, Ozaki, and Yokota]{ootomo2023dgemm}
Ootomo,~H.; Ozaki,~K.; Yokota,~R. DGEMM on integer matrix multiplication unit.
  \emph{The International Journal of High Performance Computing Applications}
  \textbf{2024}, \emph{38}, 297--313\relax
\mciteBstWouldAddEndPuncttrue
\mciteSetBstMidEndSepPunct{\mcitedefaultmidpunct}
{\mcitedefaultendpunct}{\mcitedefaultseppunct}\relax
\EndOfBibitem
\bibitem[Minamihata \latin{et~al.}(2016)Minamihata, Ozaki, Ogita, and
  Oishi]{ozaki_conference}
Minamihata,~A.; Ozaki,~K.; Ogita,~T.; Oishi,~S. Improved extraction scheme for
  accurate floating-point summation. The 35th JSST Annual Conference Inter-
  national Conference on Simulation Technology. 2016\relax
\mciteBstWouldAddEndPuncttrue
\mciteSetBstMidEndSepPunct{\mcitedefaultmidpunct}
{\mcitedefaultendpunct}{\mcitedefaultseppunct}\relax
\EndOfBibitem
\bibitem[Mukunoki \latin{et~al.}(2021)Mukunoki, Ozaki, Ogita, and
  Imamura]{10.1145/3472456.3472493}
Mukunoki,~D.; Ozaki,~K.; Ogita,~T.; Imamura,~T. Accurate Matrix Multiplication
  on Binary128 Format Accelerated by Ozaki Scheme. Proceedings of the 50th
  International Conference on Parallel Processing. New York, NY, USA,
  2021\relax
\mciteBstWouldAddEndPuncttrue
\mciteSetBstMidEndSepPunct{\mcitedefaultmidpunct}
{\mcitedefaultendpunct}{\mcitedefaultseppunct}\relax
\EndOfBibitem
\bibitem[Hida \latin{et~al.}(2001)Hida, Li, and Bailey]{Hida2001}
Hida,~Y.; Li,~X.~S.; Bailey,~D.~H. Algorithms for quad-double precision
  floating point arithmetic. Proceedings 15th IEEE Symposium on Computer
  Arithmetic. ARITH-15 2001. 2001; pp 155--162\relax
\mciteBstWouldAddEndPuncttrue
\mciteSetBstMidEndSepPunct{\mcitedefaultmidpunct}
{\mcitedefaultendpunct}{\mcitedefaultseppunct}\relax
\EndOfBibitem
\bibitem[Henry \latin{et~al.}(2019)Henry, Tang, and Heinecke]{Henry2019}
Henry,~G.; Tang,~P. T.~P.; Heinecke,~A. Leveraging the bfloat16 Artificial
  Intelligence Datatype For Higher-Precision Computations. 2019 IEEE 26th
  Symposium on Computer Arithmetic (ARITH). 2019; pp 69--76\relax
\mciteBstWouldAddEndPuncttrue
\mciteSetBstMidEndSepPunct{\mcitedefaultmidpunct}
{\mcitedefaultendpunct}{\mcitedefaultseppunct}\relax
\EndOfBibitem
\bibitem[Fousse \latin{et~al.}(2007)Fousse, Hanrot, Lef\`{e}vre, P\'{e}lissier,
  and Zimmermann]{10.1145/1236463.1236468}
Fousse,~L.; Hanrot,~G.; Lef\`{e}vre,~V.; P\'{e}lissier,~P.; Zimmermann,~P.
  MPFR: A multiple-precision binary floating-point library with correct
  rounding. \emph{ACM Trans. Math. Softw.} \textbf{2007}, \emph{33},
  13--es\relax
\mciteBstWouldAddEndPuncttrue
\mciteSetBstMidEndSepPunct{\mcitedefaultmidpunct}
{\mcitedefaultendpunct}{\mcitedefaultseppunct}\relax
\EndOfBibitem
\bibitem[Sun \latin{et~al.}(2020)Sun, Zhang, Banerjee, Bao, Barbry, Blunt,
  Bogdanov, Booth, Chen, Cui, Eriksen, Gao, Guo, Hermann, Hermes, Koh, Koval,
  Lehtola, Li, Liu, Mardirossian, McClain, Motta, Mussard, Pham, Pulkin,
  Purwanto, Robinson, Ronca, Sayfutyarova, Scheurer, Schurkus, Smith, Sun, Sun,
  Upadhyay, Wagner, Wang, White, Whitfield, Williamson, Wouters, Yang, Yu, Zhu,
  Berkelbach, Sharma, Sokolov, and Chan]{Sun2020}
Sun,~Q.; Zhang,~X.; Banerjee,~S.; Bao,~P.; Barbry,~M.; Blunt,~N.~S.;
  Bogdanov,~N.~A.; Booth,~G.~H.; Chen,~J.; Cui,~Z.-H.; Eriksen,~J.~J.; Gao,~Y.;
  Guo,~S.; Hermann,~J.; Hermes,~M.~R.; Koh,~K.; Koval,~P.; Lehtola,~S.; Li,~Z.;
  Liu,~J.; Mardirossian,~N.; McClain,~J.~D.; Motta,~M.; Mussard,~B.;
  Pham,~H.~Q.; Pulkin,~A.; Purwanto,~W.; Robinson,~P.~J.; Ronca,~E.;
  Sayfutyarova,~E.~R.; Scheurer,~M.; Schurkus,~H.~F.; Smith,~J. E.~T.; Sun,~C.;
  Sun,~S.-N.; Upadhyay,~S.; Wagner,~L.~K.; Wang,~X.; White,~A.;
  Whitfield,~J.~D.; Williamson,~M.~J.; Wouters,~S.; Yang,~J.; Yu,~J.~M.;
  Zhu,~T.; Berkelbach,~T.~C.; Sharma,~S.; Sokolov,~A.~Y.; Chan,~G. K.-L. Recent
  developments in the PySCF program package. \emph{The Journal of Chemical
  Physics} \textbf{2020}, \emph{153}, 024109\relax
\mciteBstWouldAddEndPuncttrue
\mciteSetBstMidEndSepPunct{\mcitedefaultmidpunct}
{\mcitedefaultendpunct}{\mcitedefaultseppunct}\relax
\EndOfBibitem
\bibitem[Ratcliff \latin{et~al.}(2020)Ratcliff, Dawson, Fisicaro, Caliste,
  Mohr, Degomme, Videau, Cristiglio, Stella, D'Alessandro, Goedecker, Nakajima,
  Deutsch, and Genovese]{Ratcliff2020}
Ratcliff,~L.~E.; Dawson,~W.; Fisicaro,~G.; Caliste,~D.; Mohr,~S.; Degomme,~A.;
  Videau,~B.; Cristiglio,~V.; Stella,~M.; D'Alessandro,~M.; Goedecker,~S.;
  Nakajima,~T.; Deutsch,~T.; Genovese,~L. Flexibilities of wavelets as a
  computational basis set for large-scale electronic structure calculations.
  \emph{The Journal of Chemical Physics} \textbf{2020}, \emph{152},
  194110\relax
\mciteBstWouldAddEndPuncttrue
\mciteSetBstMidEndSepPunct{\mcitedefaultmidpunct}
{\mcitedefaultendpunct}{\mcitedefaultseppunct}\relax
\EndOfBibitem
\bibitem[Willand \latin{et~al.}(2013)Willand, Kvashnin, Genovese,
  V{\'a}zquez-Mayagoitia, Deb, Sadeghi, Deutsch, and Goedecker]{Willand2013}
Willand,~A.; Kvashnin,~Y.~O.; Genovese,~L.; V{\'a}zquez-Mayagoitia,~{\'A}.;
  Deb,~A.~K.; Sadeghi,~A.; Deutsch,~T.; Goedecker,~S. Norm-conserving
  pseudopotentials with chemical accuracy compared to all-electron
  calculations. \emph{The Journal of Chemical Physics} \textbf{2013},
  \emph{138}, 104109\relax
\mciteBstWouldAddEndPuncttrue
\mciteSetBstMidEndSepPunct{\mcitedefaultmidpunct}
{\mcitedefaultendpunct}{\mcitedefaultseppunct}\relax
\EndOfBibitem
\bibitem[Mohr \latin{et~al.}(2014)Mohr, Ratcliff, Boulanger, Genovese, Caliste,
  Deutsch, and Goedecker]{Mohr2014}
Mohr,~S.; Ratcliff,~L.~E.; Boulanger,~P.; Genovese,~L.; Caliste,~D.;
  Deutsch,~T.; Goedecker,~S. Daubechies wavelets for linear scaling density
  functional theory. \emph{The Journal of Chemical Physics} \textbf{2014},
  \emph{140}, 204110\relax
\mciteBstWouldAddEndPuncttrue
\mciteSetBstMidEndSepPunct{\mcitedefaultmidpunct}
{\mcitedefaultendpunct}{\mcitedefaultseppunct}\relax
\EndOfBibitem
\bibitem[Jensen(2014)]{Jensen2014}
Jensen,~F. Unifying General and Segmented Contracted Basis Sets. Segmented
  Polarization Consistent Basis Sets. \emph{Journal of Chemical Theory and
  Computation} \textbf{2014}, \emph{10}, 1074--1085\relax
\mciteBstWouldAddEndPuncttrue
\mciteSetBstMidEndSepPunct{\mcitedefaultmidpunct}
{\mcitedefaultendpunct}{\mcitedefaultseppunct}\relax
\EndOfBibitem
\bibitem[Perdew \latin{et~al.}(1996)Perdew, Burke, and
  Ernzerhof]{10.1103/PhysRevLett.77.3865}
Perdew,~J.~P.; Burke,~K.; Ernzerhof,~M. Generalized Gradient Approximation Made
  Simple. \emph{Physical Review Letters} \textbf{1996}, \emph{77},
  3865--3868\relax
\mciteBstWouldAddEndPuncttrue
\mciteSetBstMidEndSepPunct{\mcitedefaultmidpunct}
{\mcitedefaultendpunct}{\mcitedefaultseppunct}\relax
\EndOfBibitem
\bibitem[Stephens \latin{et~al.}(1994)Stephens, Devlin, Chabalowski, and
  Frisch]{Stephens1994}
Stephens,~P.~J.; Devlin,~F.~J.; Chabalowski,~C.~F.; Frisch,~M.~J. Ab Initio
  Calculation of Vibrational Absorption and Circular Dichroism Spectra Using
  Density Functional Force Fields. \emph{The Journal of Physical Chemistry}
  \textbf{1994}, \emph{98}, 11623--11627\relax
\mciteBstWouldAddEndPuncttrue
\mciteSetBstMidEndSepPunct{\mcitedefaultmidpunct}
{\mcitedefaultendpunct}{\mcitedefaultseppunct}\relax
\EndOfBibitem
\bibitem[Pulay(1982)]{Pulay1982}
Pulay,~P. Improved SCF convergence acceleration. \emph{Journal of Computational
  Chemistry} \textbf{1982}, \emph{3}, 556--560\relax
\mciteBstWouldAddEndPuncttrue
\mciteSetBstMidEndSepPunct{\mcitedefaultmidpunct}
{\mcitedefaultendpunct}{\mcitedefaultseppunct}\relax
\EndOfBibitem
\bibitem[Nakajima \latin{et~al.}(2015)Nakajima, Katouda, Kamiya, and
  Nakatsuka]{Nakajima2015}
Nakajima,~T.; Katouda,~M.; Kamiya,~M.; Nakatsuka,~Y. NTChem: A high-performance
  software package for quantum molecular simulation. \emph{International
  Journal of Quantum Chemistry} \textbf{2015}, \emph{115}, 349--359\relax
\mciteBstWouldAddEndPuncttrue
\mciteSetBstMidEndSepPunct{\mcitedefaultmidpunct}
{\mcitedefaultendpunct}{\mcitedefaultseppunct}\relax
\EndOfBibitem
\bibitem[Dawson \latin{et~al.}(2023)Dawson, Kawashima, Ratcliff, Kamiya,
  Genovese, and Nakajima]{Dawson2023}
Dawson,~W.; Kawashima,~E.; Ratcliff,~L.~E.; Kamiya,~M.; Genovese,~L.;
  Nakajima,~T. Complexity reduction in density functional theory: Locality in
  space and energy. \emph{The Journal of Chemical Physics} \textbf{2023},
  \emph{158}, 164114\relax
\mciteBstWouldAddEndPuncttrue
\mciteSetBstMidEndSepPunct{\mcitedefaultmidpunct}
{\mcitedefaultendpunct}{\mcitedefaultseppunct}\relax
\EndOfBibitem
\bibitem[Kruchinina \latin{et~al.}(2016)Kruchinina, Rudberg, and
  Rubensson]{Kruchinina2016}
Kruchinina,~A.; Rudberg,~E.; Rubensson,~E.~H. Parameterless Stopping Criteria
  for Recursive Density Matrix Expansions. \emph{J. Chem. Theory Comput.}
  \textbf{2016}, \emph{12}, 5788--5802\relax
\mciteBstWouldAddEndPuncttrue
\mciteSetBstMidEndSepPunct{\mcitedefaultmidpunct}
{\mcitedefaultendpunct}{\mcitedefaultseppunct}\relax
\EndOfBibitem
\bibitem[Uchino(2024)]{uchino_thesis}
Uchino,~Y. Study on Reliable Algorithms for Eigenvalue and Singular Value
  Decomposition and Matrix Multiplication. Ph.D.\ thesis, Shibaura Institute of
  Technology, 2024\relax
\mciteBstWouldAddEndPuncttrue
\mciteSetBstMidEndSepPunct{\mcitedefaultmidpunct}
{\mcitedefaultendpunct}{\mcitedefaultseppunct}\relax
\EndOfBibitem
\bibitem[Bor{\v{s}}tnik \latin{et~al.}(2014)Bor{\v{s}}tnik, VandeVondele,
  Weber, and Hutter]{Borstnik2014}
Bor{\v{s}}tnik,~U.; VandeVondele,~J.; Weber,~V.; Hutter,~J. Sparse matrix
  multiplication: The distributed block-compressed sparse row library.
  \emph{Parallel Computing} \textbf{2014}, \emph{40}, 47--58\relax
\mciteBstWouldAddEndPuncttrue
\mciteSetBstMidEndSepPunct{\mcitedefaultmidpunct}
{\mcitedefaultendpunct}{\mcitedefaultseppunct}\relax
\EndOfBibitem
\bibitem[Mohr \latin{et~al.}(2017)Mohr, Dawson, Wagner, Caliste, Nakajima, and
  Genovese]{Mohr2017}
Mohr,~S.; Dawson,~W.; Wagner,~M.; Caliste,~D.; Nakajima,~T.; Genovese,~L.
  Efficient Computation of Sparse Matrix Functions for Large-Scale Electronic
  Structure Calculations: The CheSS Library. \emph{Journal of Chemical Theory
  and Computation} \textbf{2017}, \emph{13}, 4684--4698\relax
\mciteBstWouldAddEndPuncttrue
\mciteSetBstMidEndSepPunct{\mcitedefaultmidpunct}
{\mcitedefaultendpunct}{\mcitedefaultseppunct}\relax
\EndOfBibitem
\bibitem[Bock \latin{et~al.}(2018)Bock, Negre, Mniszewski, Mohd-Yusof, Aradi,
  Fattebert, Osei-Kuffuor, Germann, and Niklasson]{Bock2018}
Bock,~N.; Negre,~C. F.~A.; Mniszewski,~S.~M.; Mohd-Yusof,~J.; Aradi,~B.;
  Fattebert,~J.-L.; Osei-Kuffuor,~D.; Germann,~T.~C.; Niklasson,~A. M.~N. The
  basic matrix library (BML) for quantum chemistry. \emph{The Journal of
  Supercomputing} \textbf{2018}, \emph{74}, 6201--6219\relax
\mciteBstWouldAddEndPuncttrue
\mciteSetBstMidEndSepPunct{\mcitedefaultmidpunct}
{\mcitedefaultendpunct}{\mcitedefaultseppunct}\relax
\EndOfBibitem
\bibitem[Rubensson \latin{et~al.}(2022)Rubensson, Rudberg, Kruchinina, and
  Artemov]{Rubensson2022}
Rubensson,~E.~H.; Rudberg,~E.; Kruchinina,~A.; Artemov,~A.~G. The Chunks and
  Tasks Matrix Library. \emph{SoftwareX} \textbf{2022}, \emph{19}, 101159\relax
\mciteBstWouldAddEndPuncttrue
\mciteSetBstMidEndSepPunct{\mcitedefaultmidpunct}
{\mcitedefaultendpunct}{\mcitedefaultseppunct}\relax
\EndOfBibitem
\bibitem[Nakata(2022)]{2109.13406v2}
Nakata,~M. MPLAPACK version 2.0.1 user manual. 2022\relax
\mciteBstWouldAddEndPuncttrue
\mciteSetBstMidEndSepPunct{\mcitedefaultmidpunct}
{\mcitedefaultendpunct}{\mcitedefaultseppunct}\relax
\EndOfBibitem
\bibitem[Knizia \latin{et~al.}(2011)Knizia, Li, Simon, and Werner]{Knizia2011}
Knizia,~G.; Li,~W.; Simon,~S.; Werner,~H.-J. Determining the Numerical
  Stability of Quantum Chemistry Algorithms. \emph{Journal of Chemical Theory
  and Computation} \textbf{2011}, \emph{7}, 2387--2398\relax
\mciteBstWouldAddEndPuncttrue
\mciteSetBstMidEndSepPunct{\mcitedefaultmidpunct}
{\mcitedefaultendpunct}{\mcitedefaultseppunct}\relax
\EndOfBibitem
\bibitem[Chatelain \latin{et~al.}(2018)Chatelain, Castro, Petit, Defour,
  Bieder, and Torrent]{Chatelain2018}
Chatelain,~Y.; Castro,~P. D.~O.; Petit,~E.; Defour,~D.; Bieder,~J.; Torrent,~M.
  VeriTracer: Context-enriched tracer for floating-point arithmetic analysis.
  2018 IEEE 25th Symposium on Computer Arithmetic (ARITH). 2018; pp
  61--68\relax
\mciteBstWouldAddEndPuncttrue
\mciteSetBstMidEndSepPunct{\mcitedefaultmidpunct}
{\mcitedefaultendpunct}{\mcitedefaultseppunct}\relax
\EndOfBibitem
\bibitem[Gavini \latin{et~al.}(2023)Gavini, Baroni, Blum, Bowler, Buccheri,
  Chelikowsky, Das, Dawson, Delugas, Dogan, Draxl, Galli, Genovese, Giannozzi,
  Giantomassi, Gonze, Govoni, Gygi, Gulans, Herbert, Kokott, K{\"u}hne, Liou,
  Miyazaki, Motamarri, Nakata, Pask, Plessl, Ratcliff, Richard, Rossi, Schade,
  Scheffler, Sch{\"u}tt, Suryanarayana, Torrent, Truflandier, Windus, Xu, Yu,
  and Perez]{Gavini2023}
Gavini,~V.; Baroni,~S.; Blum,~V.; Bowler,~D.~R.; Buccheri,~A.;
  Chelikowsky,~J.~R.; Das,~S.; Dawson,~W.; Delugas,~P.; Dogan,~M.; Draxl,~C.;
  Galli,~G.; Genovese,~L.; Giannozzi,~P.; Giantomassi,~M.; Gonze,~X.;
  Govoni,~M.; Gygi,~F.; Gulans,~A.; Herbert,~J.~M.; Kokott,~S.;
  K{\"u}hne,~T.~D.; Liou,~K.-H.; Miyazaki,~T.; Motamarri,~P.; Nakata,~A.;
  Pask,~J.~E.; Plessl,~C.; Ratcliff,~L.~E.; Richard,~R.~M.; Rossi,~M.;
  Schade,~R.; Scheffler,~M.; Sch{\"u}tt,~O.; Suryanarayana,~P.; Torrent,~M.;
  Truflandier,~L.; Windus,~T.~L.; Xu,~Q.; Yu,~V. W.-Z.; Perez,~D. Roadmap on
  electronic structure codes in the exascale era. \emph{Modelling and
  Simulation in Materials Science and Engineering} \textbf{2023}, \emph{31},
  063301\relax
\mciteBstWouldAddEndPuncttrue
\mciteSetBstMidEndSepPunct{\mcitedefaultmidpunct}
{\mcitedefaultendpunct}{\mcitedefaultseppunct}\relax
\EndOfBibitem
\end{mcitethebibliography}

\end{document}